\documentclass{ptephy_v1}%%%%%% to generate preprint number with ptep logo

\def\ltap{\raisebox{-.55ex}{\rlap{$\sim$}} \raisebox{.4ex}{$<$}}

\def\lsim{\mathrel{\ltap}}

\pdfoutput=1

\preprintnumber{XXXX-XXXX} %%% %%% Insert preprint number here

\usepackage{subfig} %for getting the subfigures e.g., "Figure 1a and 1b" etc.
\newcommand{\dif}{\mathrm{d}}
\newcommand{\nvec}{\mathbf{n}}

\begin{document}

\title{Measurement of Anisotropy and Search for UHECR Sources}

%%%% To generate auto affiliation numbers please use \author{}\affil{} command

\author{O. Deligny}
\affil{IPN, CNRS-IN2P3, Univ. Paris-Sud, Universit\'{e} Paris-Saclay, Orsay, France \email{deligny@ipno.in2p3.fr}}

\author{K. Kawata}
\affil{Institute for Cosmic Ray Research, University of Tokyo, Kashiwa, Chiba, Japan \email{kawata@icrr.u-tokyo.ac.jp}}

\author{P. Tinyakov}
\affil{Service de Physique Th\'{e}orique, Universit\'{e} Libre de Bruxelles (ULB), CP225 Boulevard du Triomphe, B-1050 Bruxelles, Belgium \email{petr.tiniakov@ulb.ac.be}}

%%% To include the collaborator name... Please use the command "\collaborator"
%%% For example: \collaborator{ATLAS Collaboration}

\begin{abstract}%
Ultra-high energy cosmic rays (UHECRs) are particles, likely protons and/or
nuclei, with energies up to $10^{20}~$eV that are observed through the giant
air showers they produce in the atmosphere. These particles carry the
information on the most extreme phenomena in the Universe. At these energies,
even charged particles could be magnetically rigid enough to keep track of, or
even point directly to, the original positions of their sources on the
sky. The discovery of anisotropy of UHECRs would thus signify opening of an
entirely new window onto the Universe. With the construction and operation of
the new generation of cosmic ray experiments -- the Pierre Auger Observatory
in the Southern hemisphere and the Telescope Array in the Northern one -- the
study of these particles, the most energetic ever detected, has experienced
a jump in statistics as well as in the data quality, allowing for a much
better sensitivity in searching for their sources. In this review, we
summarize the searches for anisotropies and the efforts to identify the
sources of UHECRs which have been carried out using these new data.
\end{abstract}

\subjectindex{E41, E43, F02, F03}

\maketitle

%\tableofcontents

\section{Introduction}
\label{sec:introduction}

Since the first detection of a cosmic ray (CR) particle with energy in excess of
$10^{20}~$eV by J. Linsley at the Volcano Ranch in 1963~\cite{LinsleyPRL1963},
the origin of these particles, the most energetic produced in nature, is an
enduring problem in astroparticle physics. These UHECRs
that come to Earth have sizeable hadronic cross section: their first
interaction in the atmosphere occurs roughly at the column depth of
$100$~g~cm$^{-2}$. Given that the proton column density in the Galaxy for most
of the UHECR arrival directions is of order $\simeq 10^{-2}$~g~cm$^{-2}$, and
for the Universe as a whole even smaller -- of order $\simeq
10^{-3}$~g~cm$^{-2}$ per Gpc, the particles arriving at Earth had no prior
interactions with protons except perhaps in the source or its vicinity. Adding
the requirement of stability and limiting ourselves to known particles, it
follows then that primary particles propagating over cosmological distances
should be photons, neutrinos, protons or nuclei\footnote{A possibility has been discussed
 that the observed UHECRs could be secondary particles produced in resonant
 interactions of primary UHE neutrinos on the cosmological neutrino
 ``background''~\cite{z-burst}. This possibility is now strongly disfavored by
 the limits on the UHE photon flux, so we do not consider it
 here.}. UHECRs are detected at ground level through the extensive air 
showers they induce in the atmosphere.  
Measurements of the height of the shower maximum exclude to a large
extent the presence of photons or neutrinos so that the bulk of UHECRs detected on Earth
are protons and/or nuclei ranging from hydrogen to iron elements. Current
experimental progress on the determination of the abundance of each element
goes beyond the scope of this review, so that the chemical composition is
considered here as possibly ranging from protons to iron nuclei.

There is now evidence, mainly from the stringent upper limits on the UHE
photon and neutrino fluxes~\cite{AugerPRD2015_AugerJCAP2016_TAPRD2013}, 
that UHECRs are accelerated by electromagnetic
processes and are not the products of the decay of supermassive particles as
suggested in the top-down scenarios. While the energy spectrum and the
chemical composition measurements provide constraints helping in inferring
properties of the acceleration processes, the identification of the sources
can only be achieved by capturing in the arrival directions a pattern
suggestive in an evident way of a class of astrophysical objects. This remains
however a difficult task, mostly because of the very small value of the UHECR 
intensity (flux per steradian) at Earth -- of order 
$3\times 10^{-1}~$km$^{-2}$~sr$^{-1}$~yr$^{-1}$ above $\simeq 10~$EeV 
(1~EeV=$10^{18}~$eV) -- together
with the fact that they experience magnetic deflections during propagation. To
collect an increased influx of events, a jump in cumulated exposure as well as
improved instrumentation have been achieved in the past decade. The Pierre
Auger Observatory, located in the province of Mendoza (Argentina) and covering
3000 km$^2$, has been allowing since 2004 a scrutiny of the UHECR sky --
except in the northernmost quarter -- with a total exposure of $\simeq 6.6
\times10^4~$km$^2$~sr~yr~\cite{AugerNIM2015}. Another scrutiny, mainly of the 
northern sky, has been provided by the Telescope Array, located in Utah (USA) and
covering 700 km$^2$, operating since 2008 with a total exposure of $\simeq 8.7
\times10^3~$km$^2$~sr~yr~\cite{TANIM2012}. These latest-generation experiments have
allowed an unprecedented sensitivity in searching for anisotropies. The object
of this paper is to review the different searches performed in the last decade
in the quest to find UHECR sources by making use of the data collected at
these observatories.

The general paradigm motivating the search for anisotropies at ultra-high
energies is presented in detail in section~\ref{sec:uhecr-propagation}. The
intervening magnetic fields in the extragalactic space and in the Galaxy are
responsible for making this task difficult by generically isotropizing the
arrival directions. Worse, there are large uncertainties in the estimates of
the strength for these fields. The larger deflections are expected to be due
to the Galactic magnetic field, whose strength is estimated to be of order
several microgauss in the disk of the Galaxy as inferred from optical and
synchrotron polarization measurements and Faraday rotation measures of pulsars
and extragalactic sources. CR particles with energies in excess of $10~$EeV
have then a Larmor radius exceeding the dimensions of the
Galaxy. Roughly, deflections are expected to behave as $\simeq 3^\circ
Z (E/100~{\rm EeV})^{-1}$, with $Z$ the charge of the CRs. The highest energy
particles are thus expected to have sufficient magnetic rigidity to
approximately maintain their initial arrival directions provided that the
electric charge is not too large. Moreover, the horizon of the highest energy
particles ($\gtrsim 60~$EeV) is limited as compared to that of particles of
lower energies, because the thresholds of interactions with background
radiations filling the Universe and leading to large energy losses are then
reached. In this way, only the foreground sources are expected to
populate the observed sky maps at these energies.

The erasure of the contribution of remote sources provides a natural mechanism
to suppress the unresolved isotropic ``background'', so that UHECRs should
originate from the nearby Universe, with source distances not exceeding
100~Mpc or so. With strong nearby sources present, clusters of events could
stand out from the isotropic background. Searches for excess of close pairs of
events could reveal the clusters of events on the sky thus indicating the
presence of such sources.  Studies of the arrival directions that would be
suggestive of such discrete sources without recourse to catalogs of
astrophysical objects are summarized in section~\ref{sec:small-angle-anis}.

On the other hand, and even without compelling indications for discrete
sources, a correlation between UHECR arrival directions and the positions of a
class of astrophysical objects could reveal an anisotropy that would trace the
sources. Such a correlation would validate the prospects to study individual
sources with sufficient exposure. Searches for these cross-correlations with
different astrophysical objects and, more generally, with the matter
distribution in the nearby Universe are presented in
section~\ref{sec:large-angle-anis}.

Besides the cross-correlation approach, the non-uniform distribution of
sources and the corresponding anisotropic distribution of arrival directions
of UHECRs may be revealed by studying the decomposition of the observed flux in
spherical harmonics.  For a sufficient exposure, a dipole moment in the
angular distribution could thus be captured, which is due to the density
gradient of CRs embedded in the Galaxy. A quadrupole pattern could arise as well
if nearby sources are distributed along a plane of matter such as the
Supergalactic plane.  Studying the large-scale patterns contained in the
arrival direction distribution as a function of energy is thus another
important piece of information to decipher. Such studies are reviewed in
section~\ref{sec:multipole-expansion}.

That no prominent source has been detected so far calls into question the
understanding of UHECRs prior to the construction of the latest-generation
experiments. Magnetic deflections blur the picture to a larger extent than
anticipated. This does not preclude that sources may be identified on a
collective basis rather than on an individual one in the future, but another
jump in statistics appears necessary. Prospects for such charged-particle
astronomy are discussed in the conclusion of this review.

\section{The UHECR propagation} 
\label{sec:uhecr-propagation}

\subsection{Attenuation: the GZK paradigm}
\label{sec:gzk-paradigm}

Both protons or nuclei of ultra-high energies attenuate when propagating in
the intergalactic space due to interactions with the background radiation:
cosmic microwave background (CMB), radio and infrared backgrounds. 
These interactions occur at the center-of-mass energies accessible to accelerators, 
so the cross sections are known, or -- in the case of nuclear photodisintegration -- 
can be estimated.

Depending on the energy of the primary particle and its nature, different
processes dominate the attenuation. For protons, there are two dominant
processes: $e^+e^-$ pair production on the CMB, which is important at energies
$(1-5)$~EeV~\cite{Aloisio:2006wv}, and pion photoproduction on the CMB at energies 
of around $(40-60)$~EeV. At energies around and above $10$~EeV, with which we are mostly concerned in this paper, it is thus the second process which is important,
causing the so-called Greizen-Zatsepin-Kuzmin (GZK) effect~\cite{GZK}. The GZK 
effect leads to a cutoff in the spectrum at energies higher than $\gtrsim 50$~EeV; 
at higher energies it limits the propagation distance for protons to several 
tens of Mpc.

In the case of nuclei, the dominant energy loss results from nuclear
photodisintegration. The nucleon binding energies are of order of a few MeV with 
relatively small variation. The photodisintegration process becomes thus efficient 
when the energy of a background photon boosted in the nucleus rest frame is of the order of several MeV, the energy necessary to split off an individual nucleon, which is 
the most frequent outcome of the reaction. The relevant parameter governing the
reaction is thus the $\Gamma$ relativistic factor of nuclei. 

For light nuclei, the $\Gamma$ factor is high enough at the energies of interest 
to induce photodisintegrations on the CMB background, so that the attenuation 
of light nuclei is much faster than that of protons, qualitatively, the faster the lighter
nucleus. For heavy nuclei like iron, the $\Gamma$ factor remains below $10^{10}$ 
even at energies as high as $100$~EeV. The photodissociation occurs then on infrared
background photons with energies about an order of magnitude higher than the typical
CMB photon energy. Since these photons are less abundant than the CMB photons, the
attenuation of heavy nuclei is relatively slow and is comparable to that of protons. In 
any case, for both light and heavy nuclei, the result of a photodissociation is most often 
a nucleon and a lighter nucleus with the same $\Gamma$ factor, which in turn is subject
to further photodisintegration.

The resulting UHECR spectrum and composition have been studied in detail both
for protons~\cite{GZKprotons} as well as for heavier nuclei~\cite{GZKnuclei}, 
and specific benchmark scenarios can be obtained by numerical simulations with 
any of several existing propagation public codes~\cite{codes}.

\begin{figure}[!h]
\centering\includegraphics[width=2.9in]{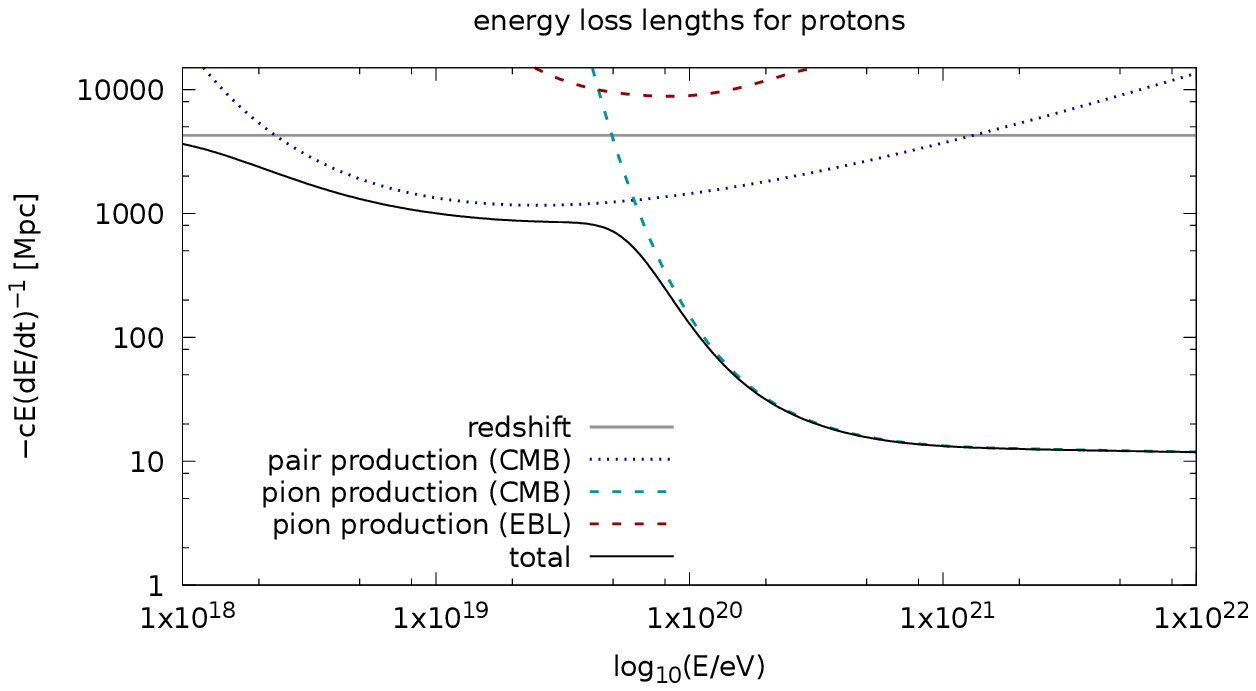}
\centering\includegraphics[width=2.9in]{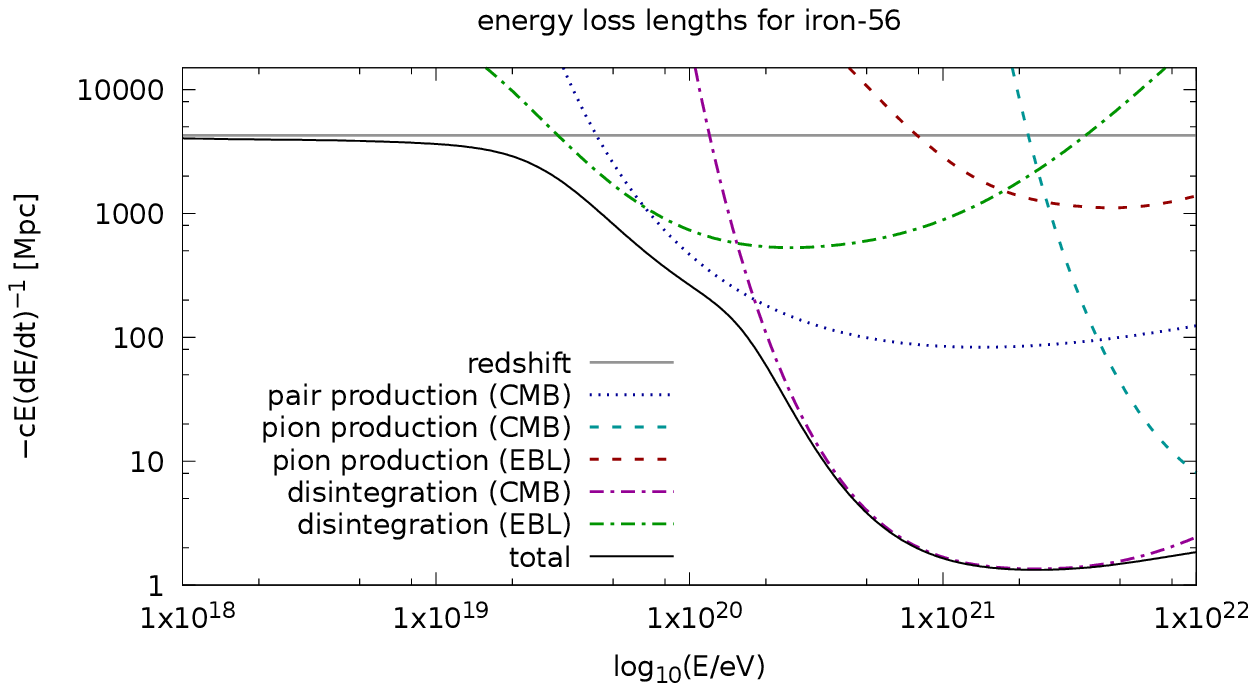}
\caption{Energy loss lengths for protons (left) and iron nuclei (right) as a 
function of energy, at redshift $z=0$, obtained using the interaction rate 
tables from~\cite{CRPropa3} and the extragalactic background light (EBL) 
model from~\cite{Gilmore2012}.}
\label{fig:energy_loss_length}
\end{figure}

The resulting energy loss lengths strongly depend on the energy of the
particles. Examples for protons and iron nuclei are shown in
Fig.~\ref{fig:energy_loss_length}, where contributions from the 
extragalactic background light and the CMB are separated. The cross sections 
for pair production can be analytically computed via the Bethe-Heitler formula, 
while those for pion photoproduction have been precisely measured in 
accelerator-based experiments and can be accurately modeled~\cite{SOPHIA}. 
In contrast, the cross sections for photo-disintegration of nuclei, especially 
for exclusive channels in which charged fragments are ejected, have only been 
measured in a few cases so that phenomenological models have to be used to 
estimate them. A comprehensive estimation of the systematic uncertainties 
affecting the UHECR propagation, mainly due to the photodisintegration cross 
sections and to the lack of knowledge of the energy spectrum and redshift 
evolution of the EBL can be found in~\cite{AlvesBatista2016}. 

Overall, the attenuation of UHECRs eliminates the contribution of distant
sources to the observed intensity on Earth. Due to this mechanism, all the
observed UHECRs should be coming from nearby sources, the higher the energy,
the smaller the size of the collection region.

\subsection{Deflections in cosmic magnetic fields}
\label{sec:mag-deflections}

Trajectories of charged particles are bent by the intervening magnetic
fields, so that the flux from a discrete source is spread over a
region of the sky, the size of which depends on the rigidity of the
particles. The absolute magnitude of deflections is such that a
particle of unit charge and energy $100~$EeV is deflected by
$0.53^\circ$ over a distance of 1~kpc in a regular field of magnitude
$1~\mu$G, and by $1.8^\circ$ over a distance of 50~Mpc in a random
field of magnitude $1$~nG and correlation length of $1$~Mpc. 

The extragalactic magnetic fields are assumed to be random and are
known quite poorly. From recent measurements of the Faraday rotations
of extragalactic sources, the strength of these fields is constrained
from above by $\sim 1$~nG for coherence lengths larger than $1$~Mpc
\cite{Kronberg1994_Pshirkov2016}. Some simulations indicate that the fields in
voids are even smaller. For instance, according to \cite{Dolag2005},
for protons of $40~$EeV propagating over distances of the order of
500~Mpc, the angular deflections should be of the order of $1^\circ$
except in directions of galaxy clusters and filaments where the
encountered magnetic fields are more intense and deflections are thus
larger. Hence, given the reduced horizon of particles at the highest
energies, the angular deflections in extragalactic space should remain
within a few degrees for the majority of nuclei, except for iron having 
both a larger electric charge and a deeper horizon.

On the other hand, coherent fields with strength up to $\simeq 3~\mu$G
are known to exist in the Galaxy within a disk of $\simeq 300~$pc
thickness, roughly described by a structure with arms with reversing
field direction between the arms and displaying variations of strength
within them~\cite{Pshirkov2011,Farrar2012}. Meanwhile,
there are uncertainties in the way the field falls off along the
direction perpendicular to the disk and in the Galactic halo. The main
features are a northerly directed poloidal component falling off
slowly with the distance from the disk, and oppositely directed
toroidal fields in the halo~\cite{Farrar2012,Pshirkov2011}. Depending
on the initial direction outside the Galaxy, deflections for protons
of $100$~EeV are of the order of $2-4^\circ$. Additional turbulent
fields with significant variations from arm to arm are also present on
correlation lengths of $\simeq 100~$pc. However, since no systematic
change in the propagation direction is expected from multiple small
deflections induced by these fields, a net root mean square deflection
is a few times smaller than the deflections induced by the
regular components~\cite{Tinyakov:2004pw,Pshirkov:2013wka}.

\section{Searches for clusterings at small and intermediate angular scales}
\label{sec:small-angle-anis}

The ultimate goal of CR astronomy is the study of the astrophysical sources
producing these particles. As emphasized in the introduction, there is 
hope for finding discrete sources at the highest energies thanks to the
GZK effect, because the isotropic background of CR arrival directions caused
by sources distributed throughout the distant Universe is suppressed. With 
dominating foreground sources in our part of the Universe, clusters of events 
could be detectable. Searches for clusterings at small and intermediate angular 
scales performed at the Pierre Auger Observatory and the Telescope Array are 
presented in this section. 

\subsection{Blind searches for over-densities}
\label{sec:blind-searches-over}

To search for over-densities of events over the exposed sky, a simple and
popular technique is to build a smoothed sky map by attributing the observed
number of events within a circular window with some specific radius to each
sampled point on the exposed sky. The probability of the observed number of 
events in each sample point is then computed from the binomial distribution
by estimating the expected number of events for an isotropic distribution 
within each circular window. A significance sky map is then derived\footnote{
 A widely-used technique to convert the observed probability into the significance 
 is based on the Li-Ma estimator~\cite{LiMa1983}, which allows a mimic of a Gaussian 
 process in an approximated way and thus allows for estimating the significance 
 from the observed and expected number of events only. Note, however, that such 
 a conversion can also be done in a direct way from Gaussian correspondence tables.}. 
Conventionally, positive (negative) significances correspond to over-densities 
(under-densities). 

However, by not specifying \textit{a priori} the targeted regions of the
sky where the excesses are searched for as well as the angular window radius
and the energy threshold, the probability/significance sky map obtained in 
this way suffers from the numerous performed trials. In a simple situation
in which each trial would be independent from every other, obtaining a 
probability as low as any specified threshold could always be reached by
increasing the number of trials. Hence, the number of trials needs to be
accounted for in the estimation of the final probability/significance 
characterizing an excess. In some sense, this final probability is the
original one ``penalized'' for the various scans performed on the parameters
intervening in the analysis. In the toy case aforementioned, the penalization
factor would just be the number of trials. In most cases of interest, each 
trial is however not independent of every other. To calculate the penalized 
probability of an apparent excess of events, Monte-Carlo simulations are then 
the relevant tool allowing for a perfect mimic of the procedure applied to the 
analyzed data set, reproducing the various correlations between each trial. 
Mock samples are generated following an isotropic distribution folded into the 
directional exposure of the considered experiment. The same number of events 
as in the actual data is generated, keeping the same energy distribution. On 
each of these mock samples, the set of parameters scanned of the actual data 
is optimized to capture the most significant excess anywhere on the sampled 
grid of the exposed sky. The final probability of the original excess (and so 
its associated significance) is then the number of samples yielding to more 
significant excess anywhere in the scanned parameter space normalized to the 
total number of generated samples. 

\begin{figure}[!h]
\centering\includegraphics[width=4in]{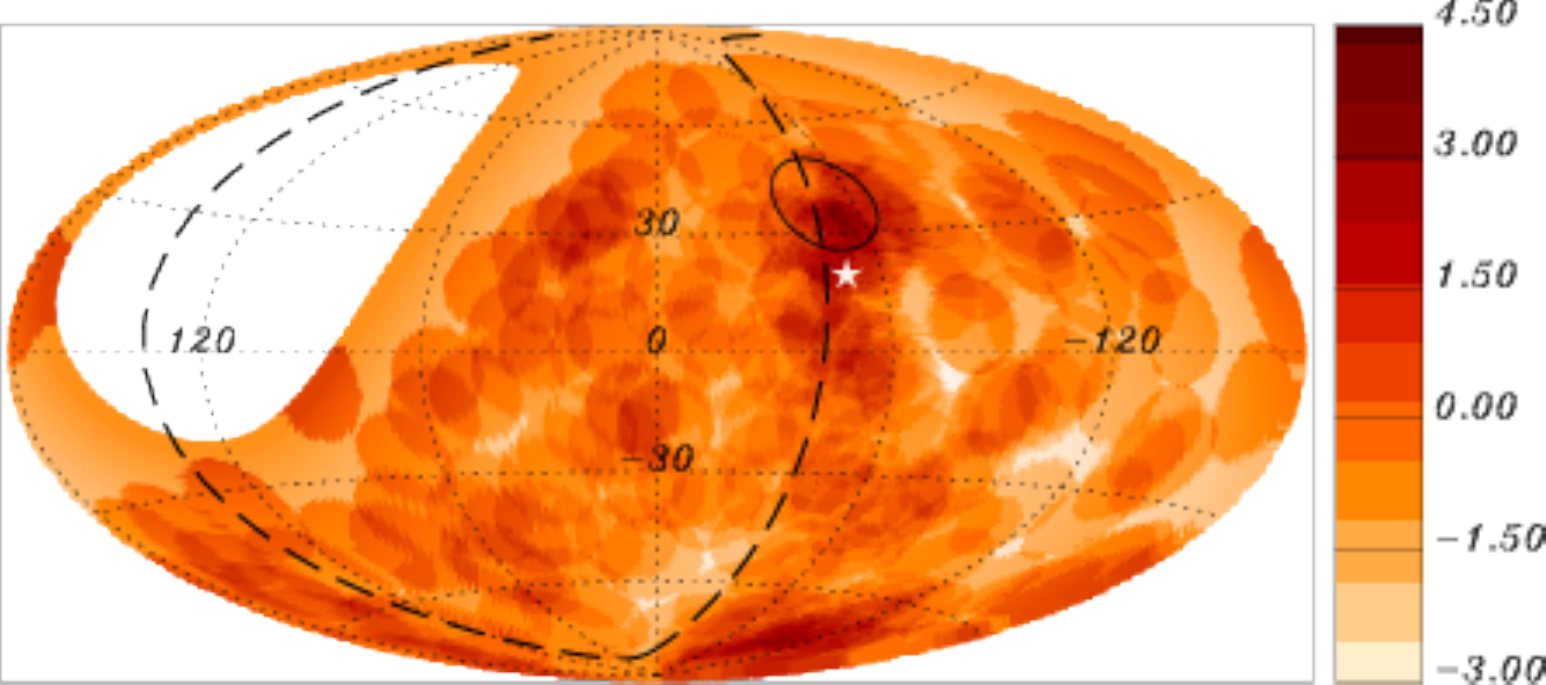}
  \caption{Map in Galactic coordinates of the Li-Ma significances of over-densities 
  in 12$^\circ$-radius windows for the events with energy in excess of 54~EeV as 
  observed at the Pierre Auger Observatory~\cite{AugerApJ2015a}.}
\label{fig:auger_hotspotsearch}
\end{figure}

The most up-to-date search for over-densities performed at the Pierre Auger
Observatory has been reported in~\cite{AugerApJ2015a}, making use of data 
recorded from January 2004 to March 2014 (corresponding to an exposure of
$\simeq 66,400$~km$^2$~sr~yr). The exposed sky was sampled using circular 
windows with radii varying from 1$^{\circ}$ up to 30$^{\circ}$ in 
1$^{\circ}$ steps, while the energy thresholds were varied from 40~EeV up 
to 80~EeV in steps of 1~EeV. The resulting (pre-trial) significance 
sky map is shown in Fig.~\ref{fig:auger_hotspotsearch} for energies in
excess of 54~EeV in 12$^\circ$-radius windows, parameters leading to the 
maximal significance. The largest departure from isotropy, indicated with
a black circle, is characterized by a pre-trial significance of 4.3$\sigma$ 
and is centered at $(\alpha,\delta)=(198^\circ,-25^\circ)$, where 14 events 
are observed against 3.23 expected from isotropy. It is close to the 
Supergalactic plane (shown as the dashed line) and centered at about 
18$^\circ$ from the direction of Centaurus A (shown as the white star). 
Once penalized for the trials, the probability of this excess is
found, however, to be as large as 69\% so that the observed  over-density 
does not provide any statistically significant evidence of anisotropy.

Similar searches have been performed on data collected at the Telescope 
Array~\cite{TAApJ2014}. One notable difference relies on the unique energy 
threshold used for this analysis, namely 57~EeV, selected from a prior 
analysis of arrival directions detected at the Pierre Auger Observatory 
which had initially led to establish an anisotropy at the 99\% confidence 
level~\cite{AugerScience2007} but which was not confirmed with 
subsequent data~\cite{AugerApJ2015a}. In addition, only oversampling radii 
of 15, 20, 25, 30, and 35 degrees were used. 

Making use of data recorded from May 2008 to May 2013 (referred to as the 
``5-year data set'' hereafter), out of the 72 selected events with zenith 
angles less than 55$^\circ$, an over-density of 19 events clustered within 
a circular window of 20$^{\circ}$ radius was observed around the equatorial 
coordinates $(\alpha,\delta)=(146.7^{\circ},43.2^{\circ})$ (near the 
Ursa Major cluster), whereas 4.5 were expected in an isotropic 
distribution (in other words, 26\% of the events were observed to be located 
within 6\% of the sky). The corresponding pre-trial significance for this 
``hotspot'' is 5.1$\sigma$, while the post-trial significance is 3.4$\sigma$.

\begin{figure}[!h]
\centering\includegraphics[width=4in]{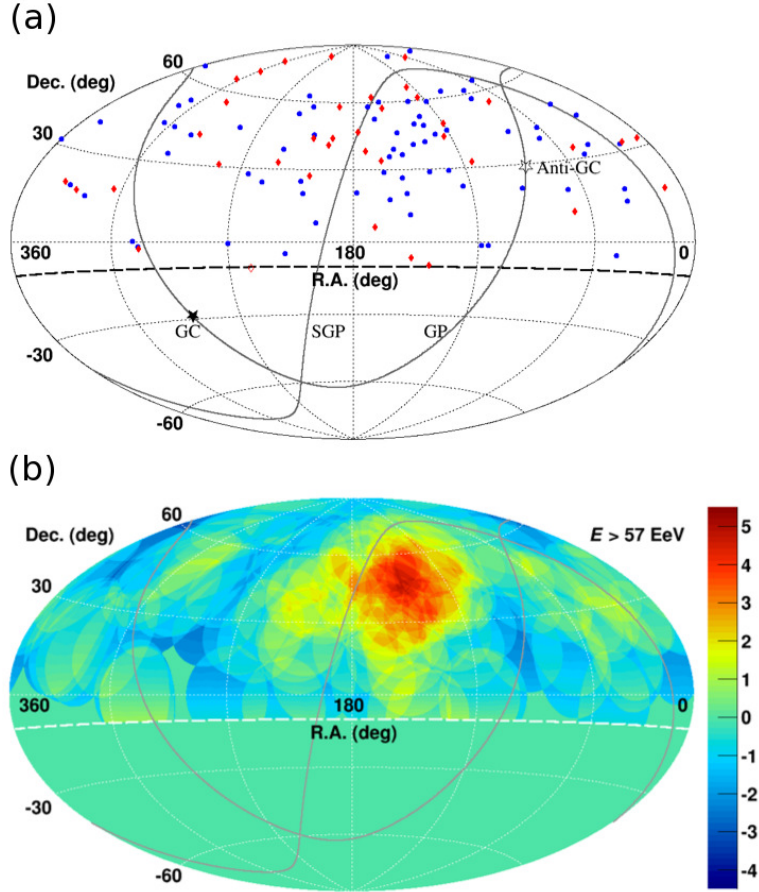}
  \caption{Aitoff projection in equatorial coordinates of the
  UHECRs detected at the Telescope Array. The solid curves stand
  for the Galactic plane (GP) and Supergalactic plane (SGP); while 
  the closed and open stars stand for the Galactic center (GC) and 
  the anti-Galactic center (Anti-GC), respectively. (a) The 
  directions of the UHECRs with $E>57$~EeV for the first 5-year
  observation time are shown as the blue points while those for the 
  6-th and 7-th observation years are shown as the red diamonds. (Note
  that one event with $\delta<-10^{\circ}$ was not included in this 
  analysis and is shown as the red open diamond). (b) Significance 
  map for the 7-year observation period using the $20^{\circ}$ 
  oversampling radius. The maximum significance is 5.1$\sigma$.}
\label{fig:hotspot}
\end{figure}

An updated analysis is now available, using data recorded between May 2008 and
May 2015 (the ``7-year data set'') \cite{Kawata:2015whq}. Out of the 109
selected events above 57~EeV and with zenith angles less than 55$^\circ$, an
over-density of 24 events is observed in a circular window of 20$^\circ$
around $(\alpha,\delta)=(148.5^{\circ},44.6^{\circ})$ while 6.88 are expected
from isotropic expectations. The position of the excess is centered
1.5$^{\circ}$ away from the one found in the previous search.  The
corresponding pre-trial and post-trial significances for this hotspot are
unchanged, namely 5.1$\sigma$ and 3.4$\sigma$, respectively. A sky map in equatorial
coordinates of the arrival directions of the 109 events with $E>57$~EeV is
shown in Fig.~\ref{fig:hotspot}~(a). The blue and red points stand for the
directions of the UHECRs for the 5-year and the latest 2-year observation
periods, respectively. In Fig.~\ref{fig:hotspot}~(b) is shown the
corresponding significance map of the excess.

As an alternative approach, the parameters that maximized the original excess 
can be fixed and used \textit{a priori} to perform an anisotropy test without
penalty factor by making use of the 6-th and 7-th year data only. In this case,
4 events are observed against 2.31 expected from an isotropic background. The 
probability of this marginal excess is estimated to be 20\%.

Overall, a definite confirmation of the signal has to await additional data with
much larger exposure. On the other hand, a survey with full-sky exposure could
provide additional information of interest in the quest to find UHECR sources.
First attempts to conduct such surveys by means of the meta-analysis of data
recorded at the Pierre Auger Observatory and the Telescope Array are currently
underway. A mapping of the entire sky for different
energy thresholds will be available in a near future.

\subsection{Autocorrelation function}
\label{sec:autoc-funct}

A clustering of CR events at a certain angular scale might first reveal itself
in the autocorrelation function of the events, which measures the cumulative
excess of event pairs separated by the given angular scale over the whole
field of view, and not necessarily localized around a single reference point
as in the approach described above in
section~\ref{sec:blind-searches-over}. This test is potentially more sensitive
than the blind search  in a situation when several small
excesses of a similar angular size are present: these contribute
coherently -- get ``stacked'' --  in the
autocorrelation function.

The autocorrelation function at a given angular scale $\psi$ can be
expressed as $(n_{\rm data}-n_{\rm bg})/n_{\rm bg}$, 
where $n_{\rm data}$ is the number
of pairs of the data events separated by the angles within the angular bin
corresponding to the scale $\psi$, and $n_{\rm bg}$ is the corresponding
number of pairs in the uniformly distributed background, usually calculated by
a Monte-Carlo simulation.  Because of the limited statistics, one usually
works not with the correlation function itself, but with the corresponding
cumulative quantity $F(\psi) = (N_{\rm data}(\psi) - N_{\rm
  bg}(\psi))/\sqrt{N_{\rm bg}(\psi)}$, with $N_{\rm data}(\psi)$ and
$N_{\rm bg}(\psi)$ being the total number of pairs separated by angles less
than $\psi$ in the data and in the simulated background, respectively. If
the data and the background are statistically identical, this quantity
fluctuates around zero. For large $N_{\rm bg}(\psi)$, positive (negative)
values of $F(\psi)$ much larger than one indicate an excess (deficit) of pairs
at a corresponding angular scale. 

The probability (more precisely, the $p$-value) of the excess, if that
is found, is determined by the Monte-Carlo simulations by counting how
often the number of pairs with the angular separation less than
$\psi$ in the simulated sets equals or exceeds $N_{\rm
  data}(\psi)$. In the limit when $1 \ll N_{\rm bg}(\psi) \ll
N_{\rm tot}$ with $N_{\rm tot}$ being the total number of events in
the data, the Poisson statistics can be used and $F(\psi)$ gives
directly the probability of the excess in equivalent Gaussian sigmas.

In practice, the angular scale $\psi$ is not fixed {\em a priori}
and is scanned over. Other parameters like the energy threshold in the
UHECR data set may also be scanned to maximize the excess. In this
case, to derive the global significance of the excess, a penalty
factor should be applied to account for the effective number of
trials, in a way similar to that described in the previous section. 

\begin{figure}[!h]
\begin{picture}(420,170)(0,0)
\put(0,0){\includegraphics[width=200pt]{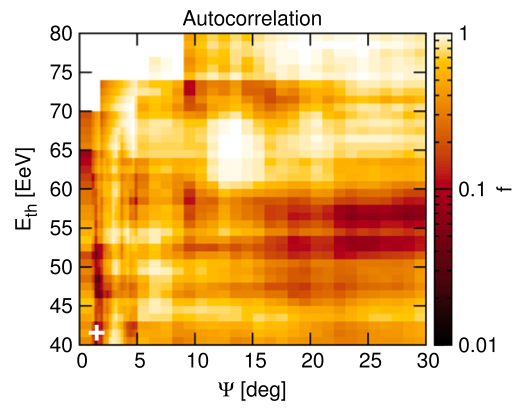}}
\put(220,10){\includegraphics[width=200pt]{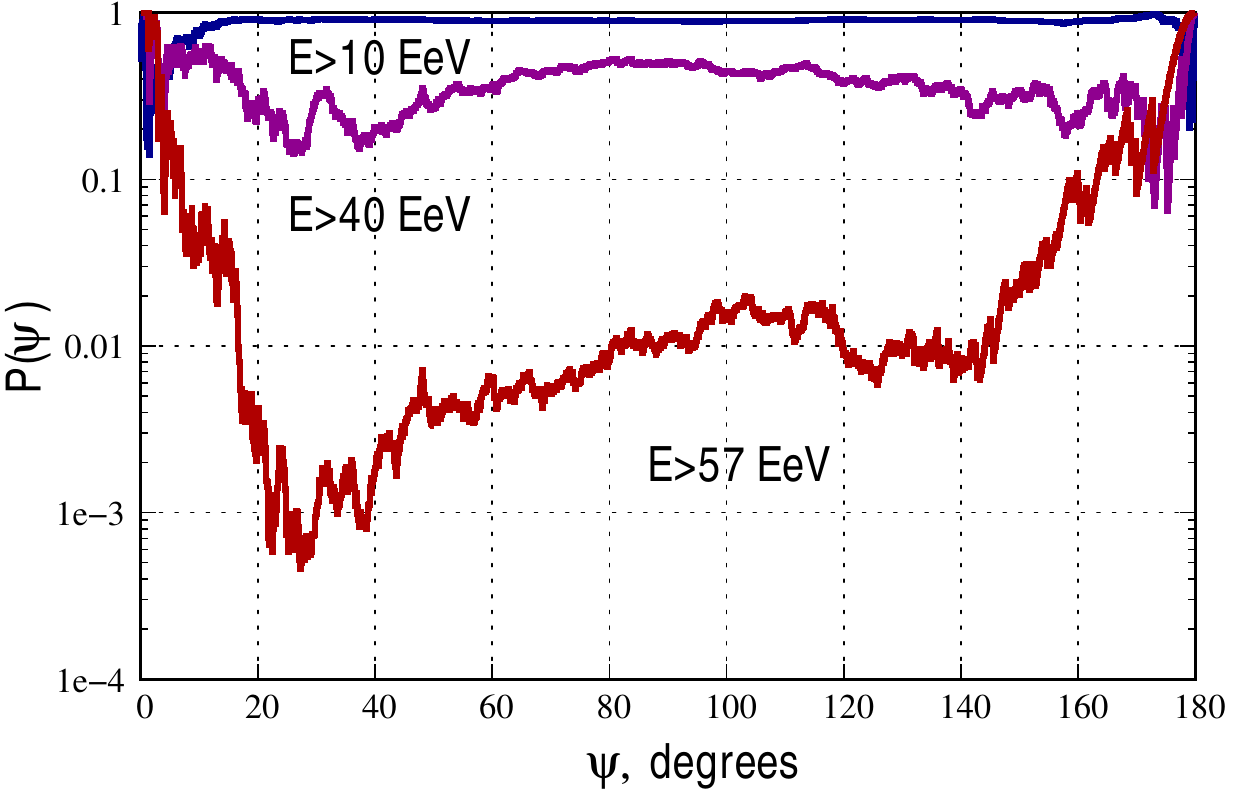}}
\end{picture}
\caption{Autocorrelation function of the UHECRs detected at the Auger
  Observatory~\cite{AugerApJ2015a} (left panel) and by the Telescope
  Array~\cite{Sagawa:2015sgk} (right panel). {\em Left panel:} the
  $p$-value (color-coded) as a function of the angular separation 
  between the pairs and the energy threshold.  {\em Right panel:} the
  $p$-value as a function of the separation angle for three energy
  thresholds as indicated on the plot. 
\label{fig:auger_auto}}
\end{figure}

The autocorrelation function of the events detected at the Pierre
Auger Observatory is presented in Fig.~\ref{fig:auger_auto} (left
panel) which shows the $p$-value (color-coded) as a function of the
angular separation $\psi$ and the energy threshold. The energy threshold
is scanned in steps of 1~EeV, while the angular separation is scanned in
steps of 0.25$^\circ$ up to 5$^\circ$ and of 1$^\circ$ for larger
separations. The largest departure from isotropic
expectations, leading to a $p$-value of 0.027, is obtained for a
separation angle of 1.5$^\circ$ and above 42~EeV, where 41 pairs are
observed against 30 expected from isotropy (shown by the cross in
Fig.~\ref{fig:auger_auto}). Once penalized for the performed scans,
it turns out that about 70\% of isotropic realizations lead to
$p$-values less than 0.027, so that no self-clustering is captured
through this analysis.

Similarly, the autocorrelation function of the events detected at the
Telescope Array~\cite{Sagawa:2015sgk} is shown in
Fig.~\ref{fig:auger_auto} (right panel). The $p$-values are shown as a
function of the separation angle $\psi$ for three energy thresholds:
$E>10$~EeV, $E>40$~EeV and $E>57$~EeV. No global significance is given
in \cite{Sagawa:2015sgk} for this analysis in view of the small deviation
from isotropy. Note, however, that largest deviations from isotropy
(the smallest $p$-values) occur in the set $E>57$~EeV at angular scales
$20^\circ-30^\circ$, consistent with the angular size of the hot spot
discussed in the previous section.

\section{Searches for correlations with nearby extragalactic matter}
\label{sec:large-angle-anis}

The sources of UHECRs being unknown, a plausible hypothesis can still
be made about their space distribution: regardless of their nature,
they must, at large scales, follow the distribution of the baryonic
matter. This fact alone may be sufficient to derive the sky
distribution of UHECRs as a function of their propagation parameters
(composition and magnetic fields, in the first place), which then may
be compared to observations to derive constraints on those
parameters. The question of sources may thus be disentangled from
other unknowns.

At scales of $\lsim 100$~Mpc, the matter distribution in the Universe
is inhomogeneous. There are large over-densities of matter
corresponding to clusters of galaxies, sheets and filaments, and
under-densities corresponding to voids. The 3D positions of the
closest of these structures -- those within several hundred cubic
megaparsecs -- are known from complete galaxy catalogs. UHECR
sources must trace this distribution to some extent. 

A key parameter in this approach is the space density of sources
$n$. In the extreme case that the sources are very rare, a few per
$(100~{\rm Mpc})^3$ or less, the source positions will appear random
on the sky despite the fact that they follow the matter distribution,
because the latter is essentially homogeneous at such large
scales. This situation is already disfavored by the existing
observations of the very high energy end of the spectrum: in the case
the distance to the closest source is exceeding several tens of Mpc, a
complete absence of the super-GZK events is expected, or at least a
very sharp cut-off.

In the opposite extreme where the sources are very numerous, they
populate the structures proportionally to the total amount of
matter. The distribution of sources will thus trace the galaxy
distribution, and so should the sky distribution of the UHECR events,
after proper accounting for the distance and propagation effects as
described in section~\ref{sec:uhecr-propagation}. This is the limit
where the UHECR flux, in principle, can be deduced from the galaxy
distribution. The nature of sources would influence the result only
slightly through different clustering properties of different types of
galaxies -- potential acceleration sites of UHECRs.

\subsection{Cross-correlation analyses} 
\label{sec:cross-corr-analys} 

If the deflections of UHECRs are not too big, their arrival directions
may show a cross correlation with the positions of the nearby
sources. A search for such a correlation is the most straightforward
way to check whether objects of a given class are sources of UHECRs
or not.

The correlation function between the UHECR events and a given catalog
of objects is calculated by counting the number of pairs
(event)-(catalog object) separated by an angular distance within the
range defined by a given angular bin, in a way similar to that
described in section~\ref{sec:small-angle-anis} for the case of the
auto-correlation function. Like in that case, for reasons of small
statistics, one usually considers the cumulative number of pairs, that
is all pairs with separation smaller than that given. The $p$-value of the
excess, if any, is determined either from the Monte-Carlo simulation
in a way similar to that described in
section~\ref{sec:small-angle-anis}, or semi-analytically by calculating
numerically the probability that a single UHECR event falls within a
given angular distance from {\em any} of the sources, and then using
the binomial distribution.

As in the case of autocorrelations, if scans over separation angle
and/or other parameters are performed, this probability needs to be
corrected for the effective number of trials, calculated again by the
Monte-Carlo simulation where synthetic data sets are generated
assuming a uniform distribution of the incident particles and
passing each set through the same search procedure as the real data.
Several analyses of this type have been performed by both Auger and TA
collaborations, which we describe below.

An important point to keep in mind is that what is actually
tested by any analysis of this type is the hypothesis that the
distribution of the UHECR arrival directions is uniform. Low
probabilities indicate that this hypothesis is false, but do not imply
by itself that the catalog objects {\em are} sources of UHECRs: the
actual sources may simply trace the distribution of the catalog
objects.

The cross-correlation tests with catalogs performed by the Auger collaboration
are described in~\cite{AugerApJ2015a} which we follow here. Several
classes of potential sources were considered: the 2MRS catalog of galaxies
\cite{Huchra:2011ii}, the Swift-BAT catalog of AGNs \cite{Baumgartner:2012qx}
and a catalog of radio galaxies with jets~\cite{vanVelzen:2012fn}. For each of
the catalogs, the scans were performed over the CR threshold energy
$40~{\rm EeV}< E_{\rm th}< 80~{\rm EeV}$, the angular scale $1^\circ < \psi< 
30^\circ$ and the catalog distance cut from 10 to 200 Mpc. In each case the
post-trial probability $P$ has been calculated. 

\begin{figure}[!h]
\begin{picture}(432,130)(0,0)
\put(0,0){\includegraphics[width=144pt]{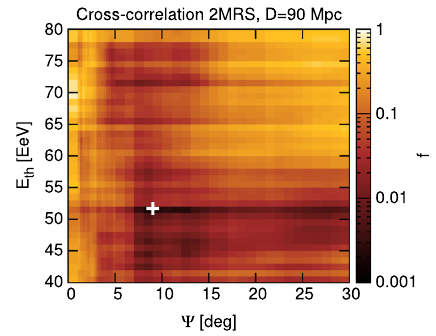}}
\put(144,0){\includegraphics[width=144pt]{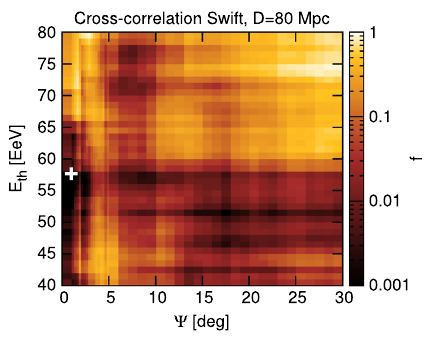}}
\put(288,0){\includegraphics[width=144pt]{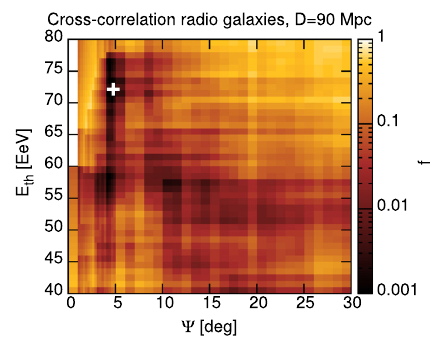}}
\end{picture}
\caption{The cross correlation of the UHECR events observed by the
  Pierre Auger Observatory with the 2MRS catalog of galaxies (left),
  the Swift-BAT catalog of AGNs (middle) and the catalog of radio
  galaxies with jets (right). Each plot shows the dependence of the
  $p$-value (color-coded) on the energy threshold in the UHECR data
  set and the angular separation. The crosses mark the best (minimum)
  $p$-values, which are $1.5\times 10^{-3}$,  $6\times 10^{-5}$ 
  and  $2\times 10^{-4}$ for the three cases from left to right,
  respectively. 
\label{fig:auger-2MRS}}
\end{figure}

Figs.~\ref{fig:auger-2MRS} show the non-penalized (pre-trial)
$p$-values for the three catalogs considered as indicated on the
plots. In each case, the distance cut in the catalog is set to give
the minimum $p$-value; these are cited in the caption. After including
the trial factors, the post-trial probabilities equal 8\%, 1\% and
1.4\% for the 2MRS galaxies, Swift-BAT sources and radio galaxies,
respectively. No significant correlations are found.

In the case of the X-ray and radio catalogs, an alternative search was also
performed where the maximum distance cut was replaced by the cut on the
minimum intrinsic luminosity $\cal L$. This cut was scanned over.  The
following post-trial probabilities were found: 1.3\% with the cut ${\cal L} >
10^{44}$~erg~s$^{-1}$ for the Swift-BAT catalog, and 11\% with the cut ${\cal
  L} > 10^{40}$~erg~s$^{-1}$ for the case of radio galaxies. It was concluded
that no significant correlations were observed.

An analogous search has been performed by the TA collaboration with the
Northern sky events \cite{Abu-Zayyad:2013vza}. The following catalogs of
objects were considered: the 3CRR catalog containing radio galaxies detected
at 178~MHz with fluxes greater than 10~Jy \cite{Laing:1983ke}, excluding the
Galactic plane $|b|<10^\circ$; the 2MRS catalog \cite{Huchra:2011ii}; the
extragalactic subset of the Swift BAT catalog \cite{Baumgartner:2012qx}
consisting of objects which were detected with a significance greater than
$4.8\sigma$ in the energy range of $14 - 195$~keV in the first 58 months of
observation by Swift BAT; the compilation of the Swift BAT AGNs detected with
at least $5\sigma$ significance in the energy range of $15 -
55$~keV in the first 60 months of observation; the 2LAC set
\cite{Fermi-LAT:2011xmf} consisting of AGNs detected with at least $4\sigma$
significance in the energy range of 100 MeV $-$ 100 GeV in the
first 24 months of observation by Fermi-LAT with the exclusion of the Galactic
plane $|b|<10^\circ$; the VCV catalog \cite{VeronCetty:2006zz} which is a
compilation of several AGN surveys. In all cases, the maximum redshift cut on
the catalog objects, the minimum energy cut on CR events, and the angular
scale of correlation were considered free parameters over which the
correlation was optimized. The post-trial probability was then calculated by
repeating the scanning procedure on the large number of isotropic Monte-Carlo
sets.

\begin{figure}[h]
\centering
\includegraphics[width=0.8\columnwidth]{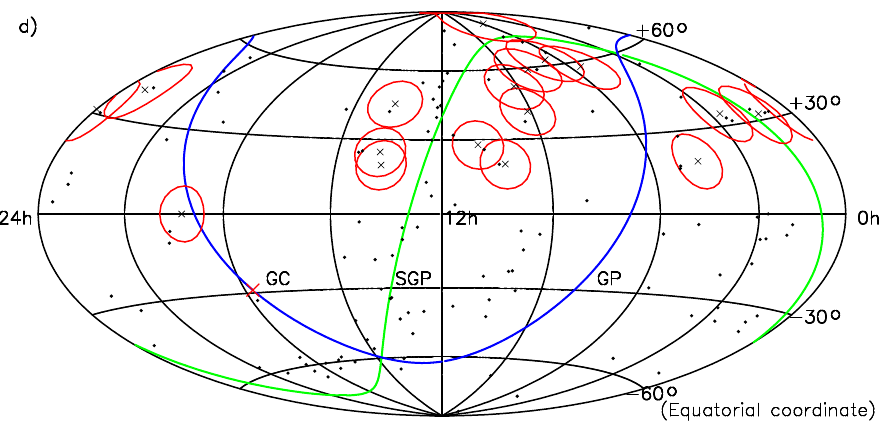}
\caption{The sky plot (Aitoff projection, equatorial coordinates) of the TA UHECR 
  events with $E>62.2$~EeV (crosses) and the objects from the Swift BAT AGN
  catalog with redshift $z<0.02$ (dots). Red circles around positions of UHECR events 
  have a radius of $10^\circ$. Blue and green lines show the Galactic and
  Supergalactic planes, respectively.
\label{fig:TA-Swift-AGN}}
\end{figure}

The strongest correlation was found with the Swift BAT AGN catalog, with the
energy threshold of $62.2$~EeV, angular scale $10^\circ$ and maximum redshift 
0.02. The sky map of the TA events and the objects corresponding to these 
cuts are shown in Fig.~\ref{fig:TA-Swift-AGN}.
The post-trial probability that such a correlation occurs as a fluctuation
over the isotropic background was estimated to be 1\% not including penalty
for searching in several catalogs. Thus, no significant correlation with the
extragalactic objects was found in the Northern sky either.

\subsection{Correlations with the Large-Scale Structure}
\label{sec:corr-with-large}

A more elaborate approach is to explicitly take into account the
catalog distances and the energy attenuation during propagation. The
drawback of this approach is, however, that it involves more unknown
parameters. First, the composition of UHECR is unknown, particularly
at the highest energies. While protons and iron nuclei attenuate in a
qualitatively similar way, the attenuation of the intermediate nuclei
is much faster, as explained in section~\ref{sec:gzk-paradigm}. As the
attenuation determines the contribution of the remote and thus 
isotropically distributed sources, the
propagation uncertainties affect primarily the overall magnitude of
the expected flux variations over the sky.

Second, the magnetic deflections may be large depending on the UHECR
composition and energy. These comprise both the regular deflections
resulting from the coherent Galactic field, and random ones which are
due to the extragalactic and random Galactic fields (see
section~\ref{sec:mag-deflections}). The latter may be characterized by a
single parameter -- a typical deflection $\theta$ (which, in general,
may depend on the direction on the sky); the former require modeling
of the coherent Galactic field which, at the moment, involves large
uncertainties.

Simplifying assumptions are needed to proceed further. In the TA
analysis (see ref.~\cite{AbuZayyad:2012hv} for details) which we
describe now, one assumes a pure proton composition. This is
consistent with the TA composition measurements \cite{Abbasi:2014sfa},
but inconsistent with the Auger composition results
\cite{Aab:2014aea}.

Once a pure proton composition is assumed, the major remaining
uncertainty is related to cosmic magnetic fields.  A further
simplifying assumption is made in the TA analysis that when the
magnetic deflections are not too large, they may be accounted for by
the random Gaussian smearing of the flux characterized by the single
angular scale $\theta$ treated as a free parameter. This smearing is
supposed to account for the finite angular resolution of the
experiment, the random deflections in the Galactic and extragalactic
turbulent fields, as well as the deflections in the regular Galactic
field. The latter part of the deflections is not random; however, for
small deflections, and for a particular type of the statistical
analyses that are less sensitive to the coherent nature of deflections
than to their overall magnitude, this approach appears a reasonable
approximation.

With these assumptions, one can model the expected sky distribution of the
UHECR flux by propagating CRs from their sources to the Earth. In the
TA analysis~\cite{AbuZayyad:2012hv}, the source distribution is assumed to
trace the galaxy distribution in the nearby Universe. The latter is obtained
from the preliminary version of the 2MRS galaxy catalog \cite{Huchra:2011ii}
which is nearly complete everywhere except in the vicinity of the Galactic
plane. The flux-limited subsample with the apparent magnitude $m\leq 12.5$ is
complete out to distances of $250$~Mpc; beyond this distance the source
distribution is taken as uniform. Each galaxy within 250 Mpc is treated as a
UHECR source of fixed intrinsic luminosity and spectrum.  Its contribution to
the total flux at energies higher than a given threshold $E_{\rm thr}$ is
calculated taking into account the distance and attenuation during 
propagation. Individual contributions are smeared with the 2D Gaussian
function of width $\theta$. The weighting correction is made to compensate
for non-equal representation of the dim and bright galaxies in a flux-limited
sample as described in detail in~\cite{Koers:2009pd}. The fraction of the
(uniform) UHECR flux coming from distances larger than 250 Mpc is calculated and
added to the total flux.

\begin{figure}[h]
\centering
\includegraphics[width=0.8\columnwidth]{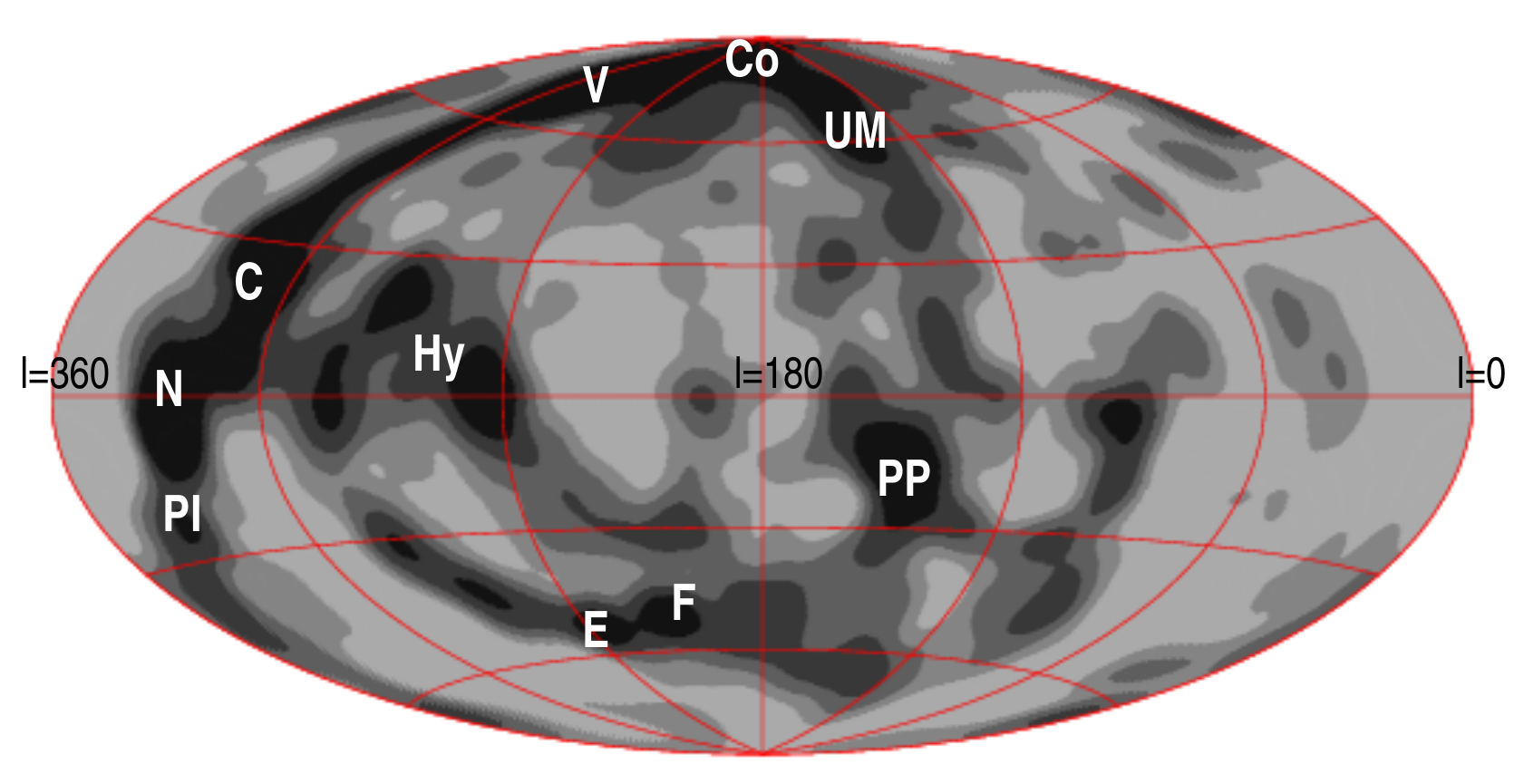}
\caption{\label{fig:LSS-canstellations} Flux distribution expected
  from sources that trace the matter in Galactic coordinates. Darker
  regions correspond to a higher flux density. The nearby structures are
  labeled as follows: C -- Centaurus supercluster (60 Mpc); Co -- Coma
  cluster (90 Mpc); E -- Eridanus cluster (30 Mpc); F -- Fornax
  cluster (20 Mpc); Hy -- Hydra supercluster (50 Mpc); N -- Norma
  supercluster (65 Mpc); PI -- Pavo-Indus supercluster (70 Mpc); PP --
  Perseus-Pisces supercluster (70 Mpc); UM -- Ursa Major supercluster
  (240 Mpc), Ursa Major North group (20 Mpc), and Ursa Major South
  group (20 Mpc); V: Virgo cluster (20 Mpc).  }
\end{figure}
An example of the flux sky map for the energy threshold $E_{\rm thr}=57$~EeV 
and the smearing angle $\theta=6^\circ$ is shown in Fig.~\ref{fig:LSS-canstellations} 
in Galactic
coordinates. Darker areas represent a larger flux density. The band boundaries
are chosen in such a way that each band integrates to the same flux. No
modulation with the experiment exposure is imposed. One may recognize the
known nearby structures; they are described in the figure caption. 

Similar maps may be constructed at different energy thresholds and
different smearing angles. The effect of changing the smearing angle
is obvious. Moving the energy threshold changes the overall contrast
of the map, which is encoded in the relative areas occupied by each
band. The bands become of equal area in the limit of zero
contrast. The higher the energy threshold, the higher the contrast due
to suppression of the contributions of remote sources.

In the TA analysis, the statistical significance of the correlation
between the expected flux and the actual distribution of events is
assessed by means of the ``flux sampling'' test~\cite{Koers:2008ba,AbuZayyad:2012hv}. 
Given a flux map $f({\bf n})$, with any set
of events (real data or Monte-Carlo generated) one may associate the
set of values of the flux map $\{ f_i\} = \{f({\bf n}_i)\}$ read off
at the positions ${\bf n}_i$ of the events. In this sense the events
sample the map, hence the name of the test.

In the case at hand one wants to know whether the two event sets --
\textit{e.g.} the data and the uniformly generated Monte-Carlo set -- are
distributed in the same way over the sky. If the two event sets are
distributed in the same way {\em on the sphere}, the two associated
sets of the flux values also have statistically equivalent
distributions: \textit{e.g.} large (or small) flux values would appear in
these two sets equally often. If one finds that this is not the case,
the {\em space} distributions of the events in the two sets must
also be different. The test is binless and does not require that the
two sets compared have the same number of events.

\begin{figure}[h]
\centering
\includegraphics[width=0.8\columnwidth]{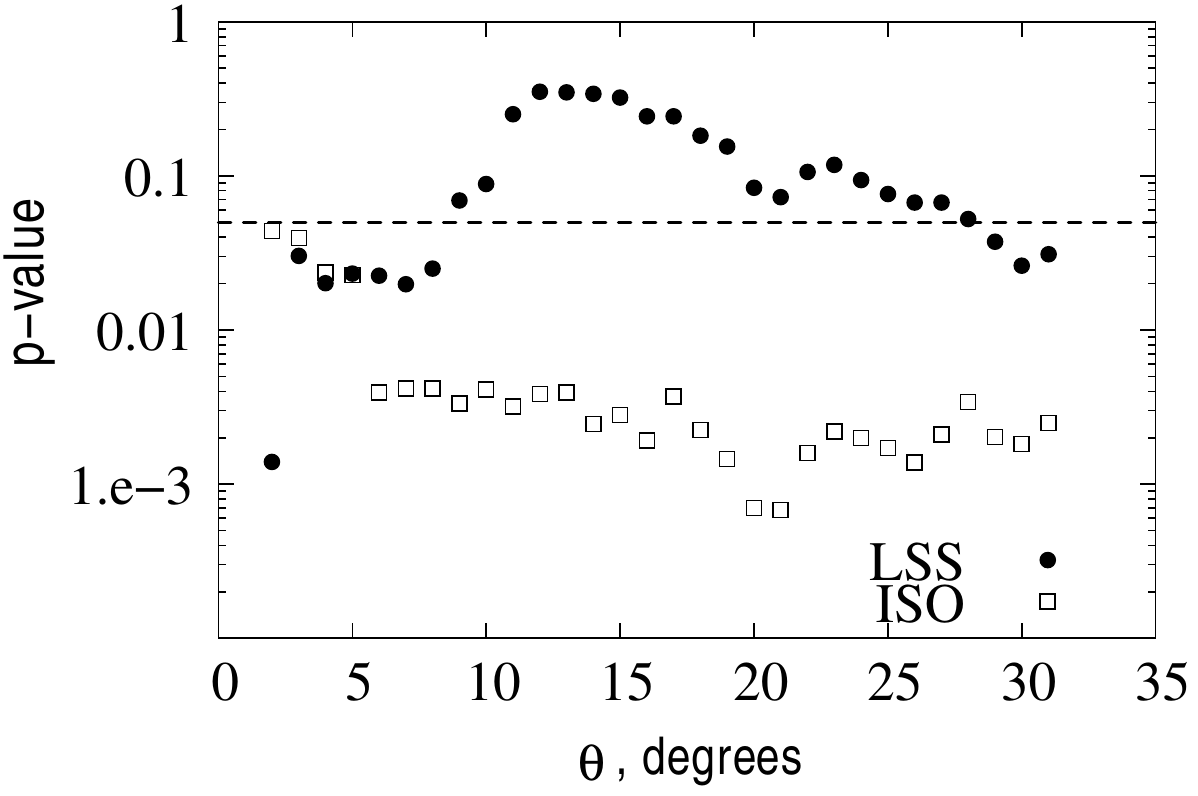}
\caption{\label{fig:LSS-pvalues} The results of a statistical test for
compatibility between the hypothesis of the isotropic UHECR distribution
(ISO, filled circles) or that following the matter distribution (LSS,
empty squares) and the TA
events with $E>57$~EeV, as a function of the typical deflection
angle. The simulation errors are smaller than the point size (not
shown on the plot). The horizontal dashed line shows the 95\% C.L. }
\end{figure}

In practice, one first generates the expected flux map with given parameters
$E_{\rm thr}$ and $\theta$. From this map one obtains the set of flux values
for the data, $\{f_i^{\rm data}\}$. Then one generates a (large) Monte-Carlo
set of events that follow the hypothesis to be tested: the uniform
distribution or following the large-scale structure (LSS) expectations. 
The map values read off at
the positions of these events give another set of flux values, $\{f_i^{\rm
  MC}\}$. The distributions $\{f_i^{\rm data}\}$ and $\{f_i^{\rm MC}\}$ are
then compared by the Kolmogorov-Smirnov test. If the $p$-value obtained is
small, the distributions $\{f_i^{\rm data}\}$ and $\{f_i^{\rm MC}\}$ are
statistically different, and so must be the distributions in arrival
directions in the two sets. The hypothesis is thus ruled out.

\begin{figure}[h]
\centering
\includegraphics[width=0.8\columnwidth]{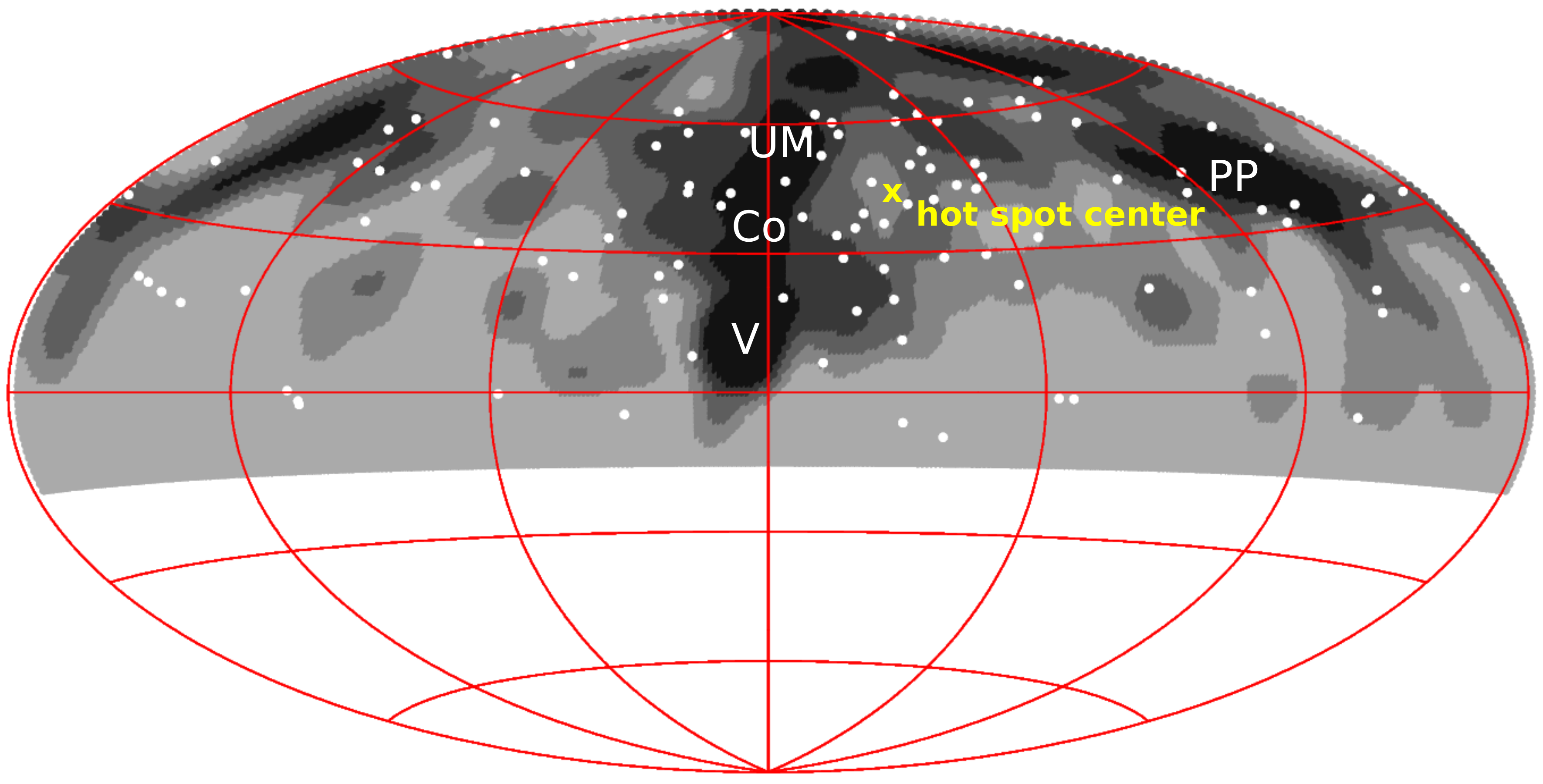}
\caption{\label{fig:LSS-skymap} The TA events with energy $E>57$~EeV
  together with the flux expected in the LSS hypothesis (sources
  follow the matter distribution) in equatorial coordinates. The cross
  shows the position of the TA hot spot. The notations for the local
  structures are the same as in Fig.~\ref{fig:LSS-canstellations}.  }
\end{figure}

In TA this analysis has been performed at three energy thresholds
$E_{\rm thr}= 10$~EeV, 40~EeV and $57$~EeV, and at smearing angles
varying from $2^\circ$ to $30^\circ$. We show in
Fig.~\ref{fig:LSS-pvalues} the results for $E_{\rm thr}= 57$~EeV
obtained with the 7 years of TA data \cite{Sagawa:2015sgk}. For each smearing
angle, two hypotheses are tested: that the distribution of events is
isotropic, and that the distribution follows the prediction of the LSS
model with a given smearing angle. The resulting $p$-values are
represented by the empty squares and filled circles, respectively. The dashed
horizontal line marks the 95\%~C.L.  The hypothesis of isotropy is
tested many times and is ruled out at about the $3\sigma$ level in most of
the individual (statistically dependent) tests. The LSS hypothesis is
in fact a different hypothesis at each smearing angle. For most,
except maybe the smallest and largest angles, this hypothesis cannot be
ruled out by this test (this does not, of course, mean that it is
true). The sky map of the TA events in equatorial coordinates,
together with the flux expected from the LSS hypothesis superimposed
with the TA exposure, are shown in Fig.~\ref{fig:LSS-skymap}. 

\section{Extracting the moments of the angular distributions}
\label{sec:multipole-expansion}

\subsection{Overview of the analysis techniques}

Large-angle structures are expected to provide the best anisotropy
fingerprints in the arrival direction distributions of UHECR events in the
energy ranges in which there are many contributing sources and/or in which the
flux from each single source is diffused over a large solid angle due to
magnetic deflections. The information on the sources is then contained in the
moments of the angular distribution of events, usually expressed in the
reciprocal space. Hence, the angular distributions of CRs are generally
characterized through the reconstructed moments describing each
corresponding angular scale. Prior to reviewing the results, a general
reminder on the formalism of the moment reconstruction is provided below.

\subsubsection{Harmonic Analysis in Right Ascension}

Extensive air shower arrays operate almost uniformly with respect to sidereal
time thanks to the rotation of the Earth : the zenith-angle-dependent shower
detection is then a function of the declination but not of the right
ascension. Thus, the most commonly used technique is the analysis in right
ascension only, through harmonic analysis of the counting rate within the
declination band defined by the detector's field of
view~\cite{Linsley1975}. Considered as a function of the right ascension only,
the flux of CRs can be decomposed in terms of a harmonic expansion:
\begin{equation}
\label{eqn:phi}
\Phi(\alpha)=a_0+\sum_{n>0}~a_n^c~\cos{n\alpha}+\sum_{n>0}~a_n^s~\sin{n\alpha}.
\end{equation}
The customary recipe to extract each harmonic coefficient makes use of
the orthogonality of the trigonometric functions. Modelling any
observed arrival direction distribution, $\overline{\Phi}(\alpha)$, as
a sum of $N$ Dirac functions over the circle,
$\overline{\Phi}(\alpha)=\sum_i \delta(\alpha,\alpha_i)$, the
coefficients can be estimated from the discrete sums:
\begin{equation}
\label{eqn:an1}
\overline{a}_n^c=\frac{2}{N} \sum_{1\leq i \leq N} \cos{n\alpha_i},\,\,\,
\overline{a}_n^s=\frac{2}{N} \sum_{1\leq i \leq N} \sin{n\alpha_i}.
\end{equation}
Here, the re-calibrated harmonic coefficients $a_n^c\equiv a_n^c/a_0$ and
$a_n^s\equiv a_n^s/a_0$ are directly considered, as it is traditionally the
case in measuring \textit{relative} anisotropies. Over-lined symbols are used
to indicate the \textit{estimator} of any quantity. The statistical properties
of the estimators $\{\overline{a}_n^c,\overline{a}_n^s\}$ can be derived from
the Poissonian nature of the sampling of $N$ points over the circle
distributed according to the underlying angular distribution
$\Phi(\alpha)$. In the case of small anisotropies (\textit{i.e.}  $|a_n^c/a_0|\ll
1$ and $|a_n^s/a_0|\ll 1$), the harmonic coefficients are recovered with an
uncertainty such that
$\sigma_n^c(\overline{a}_n^c)=\sigma_n^s(\overline{a}_n^s)=\sqrt{2/N}$. For an
isotropic realization, $\overline{a}_n^c$ and $\overline{a}_n^s$ are random
variables whose joint probability density function (p.d.f.), $p_{A_n^c,A_n^s}$, 
can be factorized in the
limit of a large number of events in terms of two Gaussian distributions whose
variances are thus $\sigma^2=2/N$. For any $n$, the joint p.d.f. of the
estimated amplitude,
$\overline{r}_n=(\overline{a}_n^{c2}+\overline{a}_n^{s2})^{1/2}$, and phase,
$\overline{\phi}_n=\arctan{(\overline{a}_n^s/\overline{a}_n^c)}$, is then
obtained through the Jacobian transformation~:
\begin{equation}
\label{eqn:jointpdf1}
p_{R_n,\Phi_n}(\overline{r}_n,\overline{\phi}_n)=\frac{\overline{r}_n}{2\pi\sigma^2}~\exp{(-\overline{r}_n^2/2\sigma^2)}.
\end{equation}
From this expression, it is straightforward to recover the Rayleigh
distribution for the p.d.f. of the amplitude, $p_{R_n}$, and the
uniform distribution between 0 and $2\pi$ for the p.d.f. of the phase,
$p_{\Phi_n}$. Overall, this formalism provides the amplitude of 
each harmonic, the corresponding phase (right ascension of the
maximum intensity), and the probability of detecting a signal due to
fluctuations of an isotropic distribution with an amplitude equal or
larger than the observed one as
$P(>\overline{r}_n)=\exp{(-N\overline{r}_n^2/4)}$. The first 
harmonic amplitude and phase, corresponding to the case $n=1$, are
of a special interest and generally draw particular attention of observers, 
since they constitute generic expectations from various models. We will
discuss the results for this harmonic obtained for EeV and trans-EeV energies 
when presenting Fig.~\ref{fig:1stharm}.

Note that the aforementioned formalism can be applied off the shelf only in
the case of an exposure that is purely uniform in the right ascension, a
condition that  is generally not fulfilled. At the \textit{sidereal} time
scale, the directional exposure of most observatories operating with high duty
cycle (\textit{e.g.} surface detector arrays) is however only moderately
non-uniform.  Different approaches are then adopted in the literature to
account for the non-uniformities. Defining $\omega(\alpha)$ as the directional
exposure integrated in declination, a widely-used and simple recipe is to
transform the observed angular distribution $\overline{\Phi}(\alpha)$ into the
one that would have been observed with a uniform directional exposure,
$\overline{\Phi}(\alpha)/\omega_{\mathrm{r}}(\alpha)$, with
$\omega_{\mathrm{r}}$ the dimensionless relative directional exposure defined
as $\omega_r(\alpha)=2\pi\omega(\alpha)/\Omega$, with $\Omega$ the total
exposure. In that way, discrete summations in equation~\ref{eqn:an1} are
changed as follows,
\begin{equation}
\label{eqn:an2}
\overline{a}_n^c=\frac{2}{\tilde{N}} \sum_{1\leq i \leq N} \frac{\cos{n\alpha_i}}{\omega_{\mathrm{r}}(\alpha_i)},\,\,\,
\overline{a}_n^s=\frac{2}{\tilde{N}} \sum_{1\leq i \leq N} \frac{\sin{n\alpha_i}}{\omega_{\mathrm{r}}(\alpha_i)},
\end{equation} 
with $\tilde{N}=\sum_i~1/\omega_{\mathrm{r}}(\alpha_i)$. The
uncertainty on the recovered coefficients then reads, still in the case
of small anisotropies~\cite{Deligny2013}, as
\begin{equation}
\label{eqn:resolution1}
\sigma_n^c=\bigg[\frac{2}{\pi\tilde{N}}\int_0^{2\pi}\frac{\mathrm{d}\alpha}{\omega_{\mathrm{r}}(\alpha)} \cos^2{n\alpha}\bigg]^{1/2},\,\,\,
\sigma_n^s= \bigg[\frac{2}{\pi\tilde{N}}\int_0^{2\pi}\frac{\mathrm{d}\alpha}{\omega_{\mathrm{r}}(\alpha)} \sin^2{n\alpha}\bigg]^{1/2}.
\end{equation}
For variations of $\omega_{\mathrm{r}}(\alpha)$ of a few percent,
these expressions are accurately approximated by
$\sigma_n^c=\sigma_n^s=\sqrt{2/\tilde{N}}$, so that in such cases, the
p.d.f. of the amplitude remains a Rayleigh distribution with the parameter
$\sqrt{2/\tilde{N}}$, while the probability of detecting a signal due
to fluctuations of an isotropic distribution with an amplitude equal
or larger than the observed one as
$P(>\overline{r}_n)=\exp{(-\tilde{N} \overline{r}_n^2/4)}$.

\subsubsection{Multipole Expansion in Right Ascension and Declination}

In general, and in contrast to the simplified approach presented in
the last subsection, the flux of CRs $\Phi(\nvec)$ can depend on both
the right ascension and the declination and thus be decomposed in
terms of a multipolar expansion in the spherical harmonics $Y_{\ell
  m}(\nvec)$:
\begin{equation}
\label{eqn:almexpansion}
\Phi(\nvec)=\sum_{\ell\geq0}\sum_{m=-\ell}^\ell a_{\ell m}Y_{\ell m}(\nvec).
\end{equation}
Non-zero amplitudes in the $\ell$ modes arise from variations of the
flux on an angular scale ${\simeq}1/\ell$ radians. With full-sky but
non-uniform coverage, the customary recipe for decoupling directional
exposure effects from anisotropy ones consists in defining the
recovered coefficients as~\cite{Sommers2001}
\begin{equation}
\label{eqn:est_alm}
\bar{a}_{\ell m}=\int_{4\pi}\frac{\dif\nvec}{\omega(\nvec)}\frac{\dif N(\nvec)}{\dif\nvec}Y_{\ell m}(\nvec),
\end{equation}
with $\omega(\nvec)$ the directional exposure providing the time-integrated 
surface of the experiment to each direction of the sky, and $\dif N(\nvec)/\dif\nvec$
the observed angular distribution. Modeling this latter distribution as a sum of
Dirac functions in the directions of each event $\dif N(\nvec)/\dif\nvec=\sum_i\delta(\nvec,\nvec_i)$, the coefficients can be estimated in practice as
\begin{equation}
\label{eqn:est_alm2}
\bar{a}_{\ell m}=\frac{4\pi f_1}{\Omega}\sum_{i=1}^N\frac{Y_{\ell m}(\nvec_i)}{\omega_\mathrm{r}(\nvec_i)},
\end{equation}
where $\omega_\mathrm{r}(\nvec)$ is the relative directional
exposure function normalized here to unity at its maximum, and $f_1=\int\dif\nvec~\omega_{\mathrm{r}}/4\pi$ is the covered fraction
of the sky. In the case of
small anisotropies, the uncertainty $\sigma_{\ell m}$ on each $a_{\ell
  m}$ multipole reflects the Poisson fluctuations induced by the
finite number of events:
\begin{equation}
\label{eqn:rms_alm}
\sigma_{\ell m}=\left[\frac{4\pi f_1N}{\Omega^2}\int_{4\pi}\frac{\dif\nvec}{\omega_{\mathrm{r}}(\nvec)}~Y_{\ell m}^2(\nvec)\right]^{1/2}.
\end{equation}

However, with ground-based observatories, coverage of the full sky is not
presently possible with a single experiment. The partial-sky coverage of
ground-based observatories prevents the multipolar moments $a_{\ell m}$ to be
recovered in the direct way just presented. This is because the solid angle on
the sky where the exposure is zero prevents one from making use of the
completeness relation of the spherical harmonics. Indirect procedures have to
be used, one of them consisting in considering first the ``pseudo''-multipolar
moments
\begin{equation}
\tilde{a}_{\ell m} = \int \dif\mathbf{n}~\omega(\mathbf{n})\Phi(\mathbf{n})Y_{\ell m}(\mathbf{n}),
\end{equation}
and then the system of linear equations relating these pseudo
moments to the real ones:
\begin{equation}
\tilde{a}_{\ell m} = \sum_{\ell'\geq0}\sum_{m'=-\ell}^\ell a_{\ell' m'}\int \dif\mathbf{n}~\omega(\mathbf{n})Y_{\ell m}(\mathbf{n})Y_{\ell' m'}(\mathbf{n})\equiv \sum_{\ell'\geq0}\sum_{m'=-\ell}^\ell [K]_{\ell m\ell' m'}~a_{\ell' m'}.
\end{equation}
Formally, the coefficients $a_{\ell m}$ appear related to  $\tilde{a}_{\ell
  m}$  through a convolution. The matrix $K$, which imprints the
interferences between modes induced by the non-uniform and partial coverage of
the sky, is entirely determined by the directional exposure function
$\omega(\nvec)$. Assuming a bound $\ell_{\mathrm{max}}$ beyond which $a_{\ell
  m}=0$, these relations can be inverted allowing recovery of the
moments $a_{\ell m}$. However, the obtained uncertainty on each moment does
not behave as expressed in equation~\ref{eqn:rms_alm}. In contrast, the
uncertainty $\sigma_{\ell m}$ on the recovered $a_{\ell m}$ coefficients
behaves as~\cite{Billoir2008}
\begin{equation}
\label{eqn:rms_alm_bis}
\sigma_{\ell m} \simeq \left[\frac{N}{\Omega}[K^{-1}_{\ell_{\mathrm{max}}}]_{\ell m\ell m}\right]^{1/2}.
\end{equation}
The inverse matrix $K^{-1}$ is here indexed by the bound $\ell_{\mathrm{max}}$,
because of the dependence of the matrix coefficients on this parameter. In
numbers, it turns out that $\sigma_{\ell m}$ increases exponentially with
$\ell_{\mathrm{max}}$. This dependence is nothing else but the mathematical
translation of it being impossible to know the angular distribution of CRs in
the uncovered region of the sky. In most of the practical cases reviewed in the
next subsections, the small values of the energy-dependent $a_{\ell m}$
coefficients combined with the available statistics in the different energy
ranges do not allow for an estimation of the individual coefficients with a
relevant resolution as soon as $\ell_{\mathrm{max}}>2$.

From the estimation of the spherical harmonic coefficients, a more geometric 
and more intuitive representation of the dipole and quadrupole moments is
generally used:
\begin{equation}
\label{eqn:phidipquad}
\Phi(\nvec)=\frac{\Phi_0}{4\pi}\left(1+r\,\mathbf{d}\cdot\nvec+\lambda_+(\mathbf{q}_+\cdot\nvec)^2+\lambda_0(\mathbf{q}_0\cdot\nvec)^2+\lambda_-(\mathbf{q}_-\cdot\nvec)^2+\ldots\right).
\end{equation}
In this picture, the dipole moment is thus characterized by a vector, whose amplitude
$r$ and two angles of the unit vector $\mathbf{d}$ are related to the $a_{1m}$
coefficients through\footnote{Note that the angles are expressed here in a horizontal 
coordinate system, such as equatorial or Galactic systems.}
\begin{equation}
\label{eq:dip_param}
r=\frac{\sqrt{3}}{a_{00}}\left[a_{10}^2+a_{11}^2+a_{1-1}^2\right]^{1/2},\quad \delta_d=\arcsin{(\sqrt{3}a_{10}/a_{00})},\quad \alpha_d=\arctan{(a_{1-1}/a_{11})}.
\end{equation}
The amplitude $r$ corresponds to the anisotropy contrast of a dipolar flux. The
quadrupole, on the other hand, is characterized by the amplitudes 
$(\lambda_+,\lambda_0,\lambda_-)$ and unit vectors 
$(\mathbf{q}_+,\mathbf{q}_0,\mathbf{q}_-)$ which are the eigenvalues 
and eigenvectors of a second order, traceless and symmetric tensor $\mathbf{Q}$
whose five independent components are related to the $a_{2m}$ coefficients
through
\begin{equation}
Q_{xx} =  \frac{\sqrt{5}}{a_{00}}\,\left(\sqrt{3}a_{22}-a_{20}\right),\quad
Q_{xy} =  \frac{\sqrt{15}}{a_{00}}\,a_{2-2},\quad
Q_{xz} = -\frac{\sqrt{15}}{a_{00}}\,a_{21},\nonumber 
\end{equation}
\begin{equation}
\label{eq:quad_tensor}
Q_{yy} =  \frac{\sqrt{5}}{a_{00}}\,\left(\sqrt{3}a_{22}+a_{20}\right),\quad
Q_{yz} = -\frac{\sqrt{15}}{a_{00}}\,a_{2-1}.
\end{equation}
The eigenvalues are ranked from the largest to the smallest one and assigned 
to the vectors $(\mathbf{q}_+,\mathbf{q}_0,\mathbf{q}_-)$ that form the 
principal axes coordinate system. The traceless condition of the quadrupole tensor
$\mathbf{Q}$ forces the relation $\lambda_++\lambda_0+\lambda_-=0$ to be 
satisfied, so that only two of these amplitudes are independent. Hence, two 
diagnostic parameters are used to characterize a quadrupole anisotropy: the 
quadrupole magnitude that takes on the value $\lambda_+$, and the anisotropy 
contrast of a quadrupolar flux $\beta=(\lambda_+-\lambda_-)/(2+\lambda_++\lambda_-)$. The orientation 
of the quadrupole is then described by the three Euler angles determined from 
the eigenvectors corresponding to each of the principal axes and characterizing the
orientation of these principal axes with respect to some reference coordinate system.

\subsection{Harmonic/Multipolar analyses from single experiments}

\begin{figure}[!h]
\centering\includegraphics[width=2.9in]{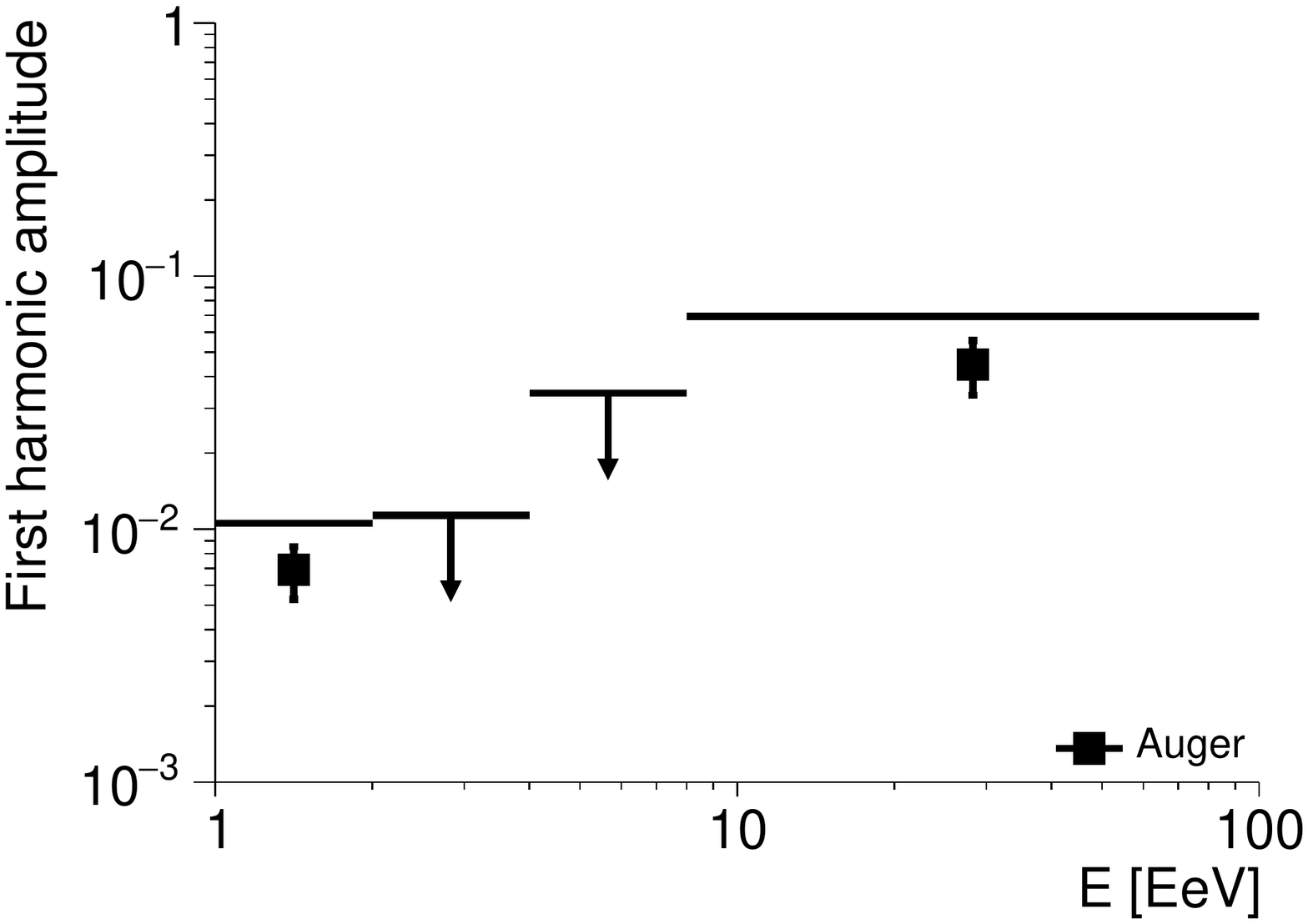}
\centering\includegraphics[width=2.9in]{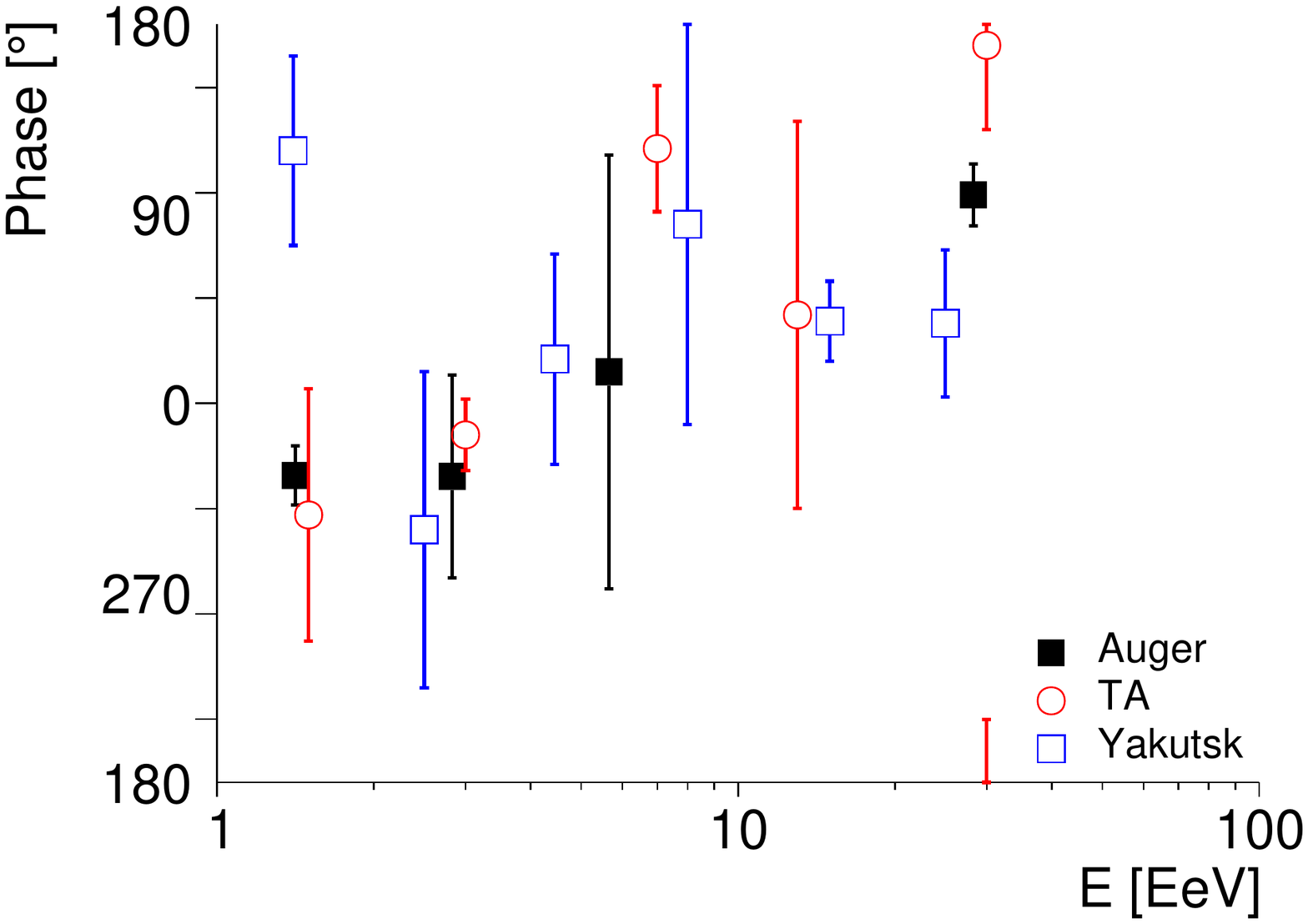}
\caption{First harmonic coefficients in right ascension above 1~EeV. Left:
  upper limits to the amplitude at the 99\% C.L. as derived at the Pierre
  Auger Observatory~\cite{AlSamarai2015}. Amplitudes are also reported (black
  squares) in the two energy bins when the corresponding $p-$value expected
  from isotropy is below $10^{-3}$. Right: Corresponding phases as obtained at
  the Pierre Auger Observatory (filled black squares)~\cite{AlSamarai2015}, at
  the Telescope Array (open red circles)~\cite{UHECR2012}, and at the Yakutsk
  array (open blue squares)~\cite{Yakutsk2001}.}
\label{fig:1stharm}
\end{figure}

Scrutiny of the large-scale distribution of arrival directions of UHECRs
provides important information in the EeV energy range. A time-honored picture
is that the ankle is a feature in the energy spectrum that is marking the
transition between Galactic and extragalactic CRs~\cite{Linsley1963}. In the
energy range just below the ankle energy, the propagation regime between the
diffusive particle transport and the deterministic flow of particles is
expected to induce large-scale anisotropies shaped by the distribution of
sources in the disk of the Galaxy and the structure of the coherent Galactic
magnetic field. On the other hand, the eventual anisotropies of
EeV-extragalactic CRs should not reflect the geometry of the Galaxy. These
generic benchmark scenarios provide some signatures that should help in
establishing the energy at which the flux of extragalactic CRs starts to
dominate the energy spectrum. Answering this old-standing question would
constitute an important step forward towards understanding the origin of
UHECRs.

Measurements of amplitudes $\overline{r}_1$ and phases 
$\overline{\phi}_1$ of the first harmonic in the right
ascension at EeV and trans-EeV energies are shown in
Fig.~\ref{fig:1stharm}. The strongest constraints on the amplitudes
are currently provided by the Pierre Auger
Observatory~\cite{AlSamarai2015}.  As in none of the energy bins are
the $p$-values for the amplitudes at the level of discovery, the upper
limits, at the 99\% C.L., are shown in the left panel. Amplitudes are also
shown in the two energy bins where the $p$-values are
$1.5\times10^{-4}$ ($1<E/\mathrm{EeV}\leq2$) and $6.4\times10^{-5}$
($E>8~$EeV)~\cite{AugerApJ2015b}. Together with these relatively low
$p$-values, an apparent consistency in the phases as measured at the
Pierre Auger Observatory~\cite{AlSamarai2015}, at the Telescope
Array~\cite{UHECR2012} and at the Yakutsk experiment from an older 
analysis~\cite{Yakutsk2001} is observed, even though the
significances of the corresponding amplitudes are relatively small. 
Note that the uncertainties on the phases are estimated as $\sqrt{2/N}/\overline{r}_1$, so that these uncertainties do not reflect 
the cumulated statistics only. As already pointed out by Linsley a long time 
ago, the observed consistency is potentially
indicative of a real underlying anisotropy, because a consistency of
the phase measurements in ordered energy intervals is indeed expected
to be revealed with a smaller number of events than needed to detect
the amplitude with high statistical
significance~\cite{Edge1978}. Interestingly, the phases derived from
experiments operating mainly in the PeV$-$EeV energy range show a
consistent tendency to align in the right ascension of the Galactic
center. In this regard, the change of phase observed above $1~$EeV
towards, roughly, the opposite of the one at energies below $1~$EeV
provides interesting information.

\begin{figure}[!h]
\centering\includegraphics[width=3.0in]{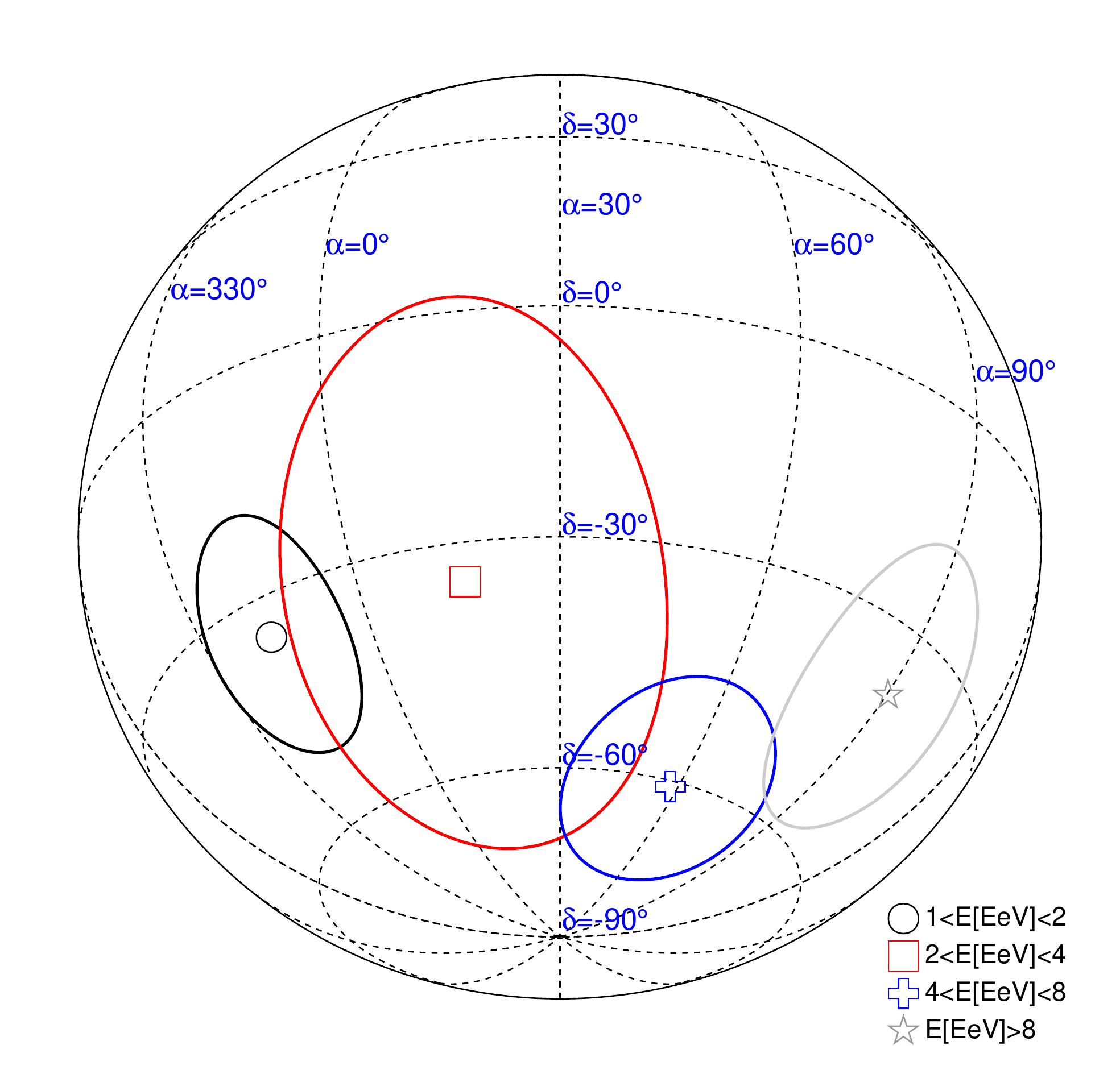}
\caption{Reconstructed declination and right-ascension of the dipole
  moment at the Pierre Auger Observatory, as a function of the energy,
  in orthographic projection~\cite{DeAlmeida2013}.}
\label{fig:dipole_directions_auger}
\end{figure}

To further characterize the EeV and trans-EeV angular distributions of
CRs, a thorough search for large-scale anisotropies in terms of dipole
and quadrupole moments was conducted by the Pierre Auger
Collaboration~\cite{AugerApJS2012}. Assuming that the anisotropic
component of the angular distributions is dominated by pure dipole or
dipole and quadrupole patterns, searches for significant moments were
performed. Within the statistical uncertainties, no strong evidence
of any significant amplitude could be captured. For a pure dipolar
flux, the reconstructed directions are shown in orthographic
projection in Fig.~\ref{fig:dipole_directions_auger} with the associated
uncertainties, as a function of the energy. The same change of phase
in right ascension as in the case of the first harmonic analysis is
observed. In addition, the reconstructed declinations are observed to
be in the equatorial southern hemisphere.

\begin{figure}[!h]
\centering\includegraphics[width=2.9in]{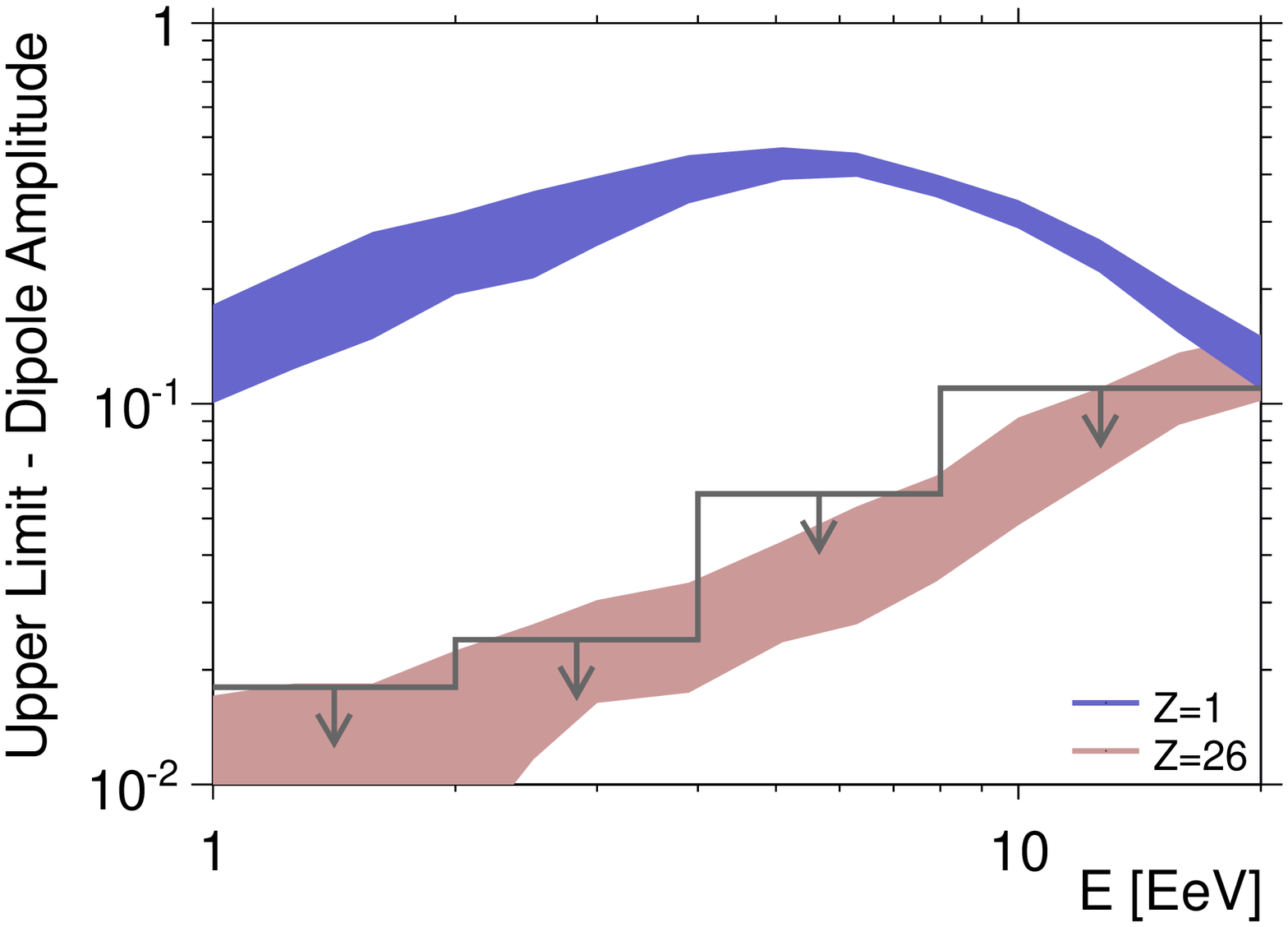}
\centering\includegraphics[width=2.9in]{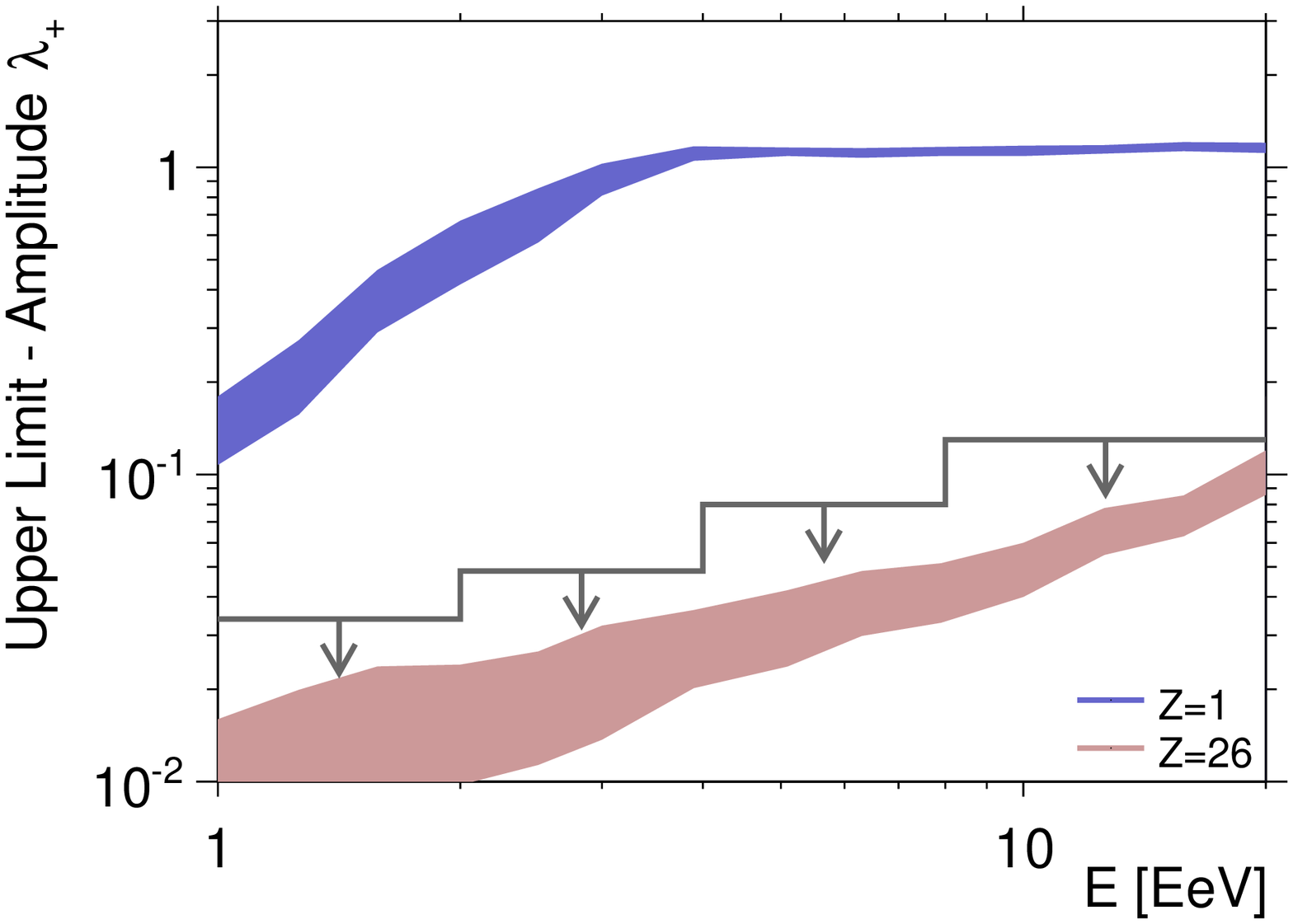}
\caption{99\% C.L. upper limits on dipole and quadrupole amplitudes as a
  function of  energy, as obtained from Auger
  data~\cite{DeAlmeida2013}. Some generic anisotropy expectations from
  stationary Galactic sources distributed in the disk are also shown, for two
  distinct assumptions on the CR composition.}
\label{fig:upplim_dipquad_auger}
\end{figure}

The obtained upper limits on dipole and quadrupole amplitudes, derived
at 99\% C.L., are shown in Fig.~\ref{fig:upplim_dipquad_auger}. The dipole
amplitude plotted in the left panel corresponds to the quantity $r$ defined 
in equation~\ref{eq:dip_param}, characterizing the anisotropy contrast
of dipolar-shaped flux. The quadrupole amplitude $\lambda_+$ plotted in the 
right panel provides the magnitude of a similar excesses at antipodal points in 
the sky. It corresponds to the greatest eigenvalue of the quadrupole tensor
presented in equation~\ref{eq:quad_tensor}. 
The bounds on the dipole amplitudes as a function of energy are shown in
the left panel along with generic estimates of the dipole amplitudes
expected from stationary Galactic sources distributed in the disk
considering two extreme cases of single primaries: protons and iron
nuclei. Both the strength and the structure of the magnetic field in
the Galaxy, known only approximately, play a crucial role in the
propagation of CRs and thus in the predictions of their arrival
directions. Best up-to-date models can be found
in~\cite{Pshirkov2011,Farrar2012}. As an illustrative case, the
bisymmetric spiral structure model with anti-symmetric halo with
respect to the Galactic plane described in~\cite{Pshirkov2011} was
considered in this study, on top of a turbulent field generated
according to a Kolmogorov power spectrum. This example is an
illustration of the potential power of these observational limits on
the dipole anisotropy to exclude the hypothesis that a light component
of EeV-CRs comes from stationary sources densely distributed in the
Galactic disk and emitting in all directions. Furthermore, assuming
that the angular distribution is modulated by a dipole and a
quadrupole, the 99\% C.L. upper bounds on the quadrupole amplitude
corresponding to the principal axis excess that could result from
fluctuations of an isotropic distribution are shown in the right panel
together with expectations considering the same astrophysical
scenario. The same conclusion holds.

\begin{figure}[!h]
\centering\includegraphics[width=2.9in]{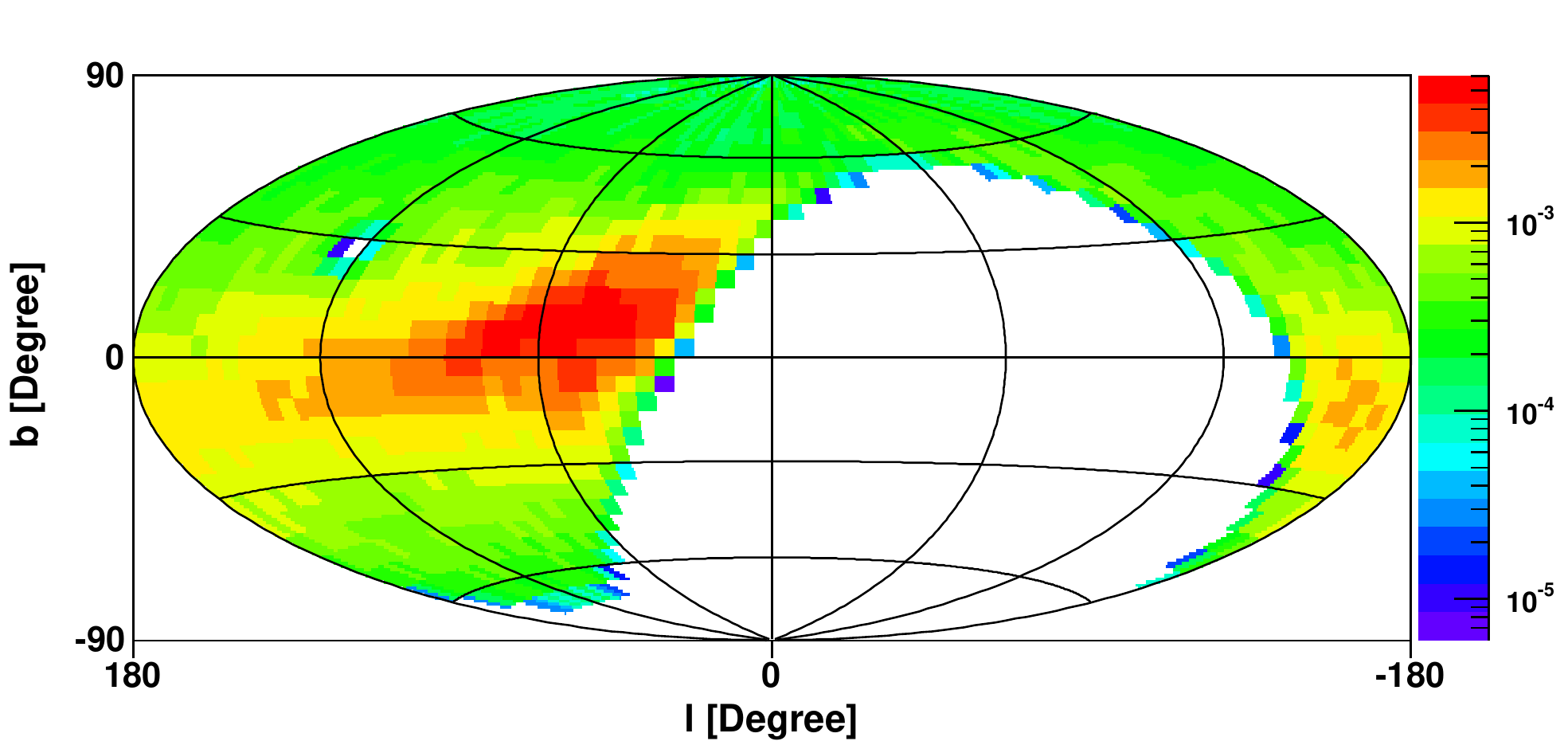}
\centering\includegraphics[width=2.9in]{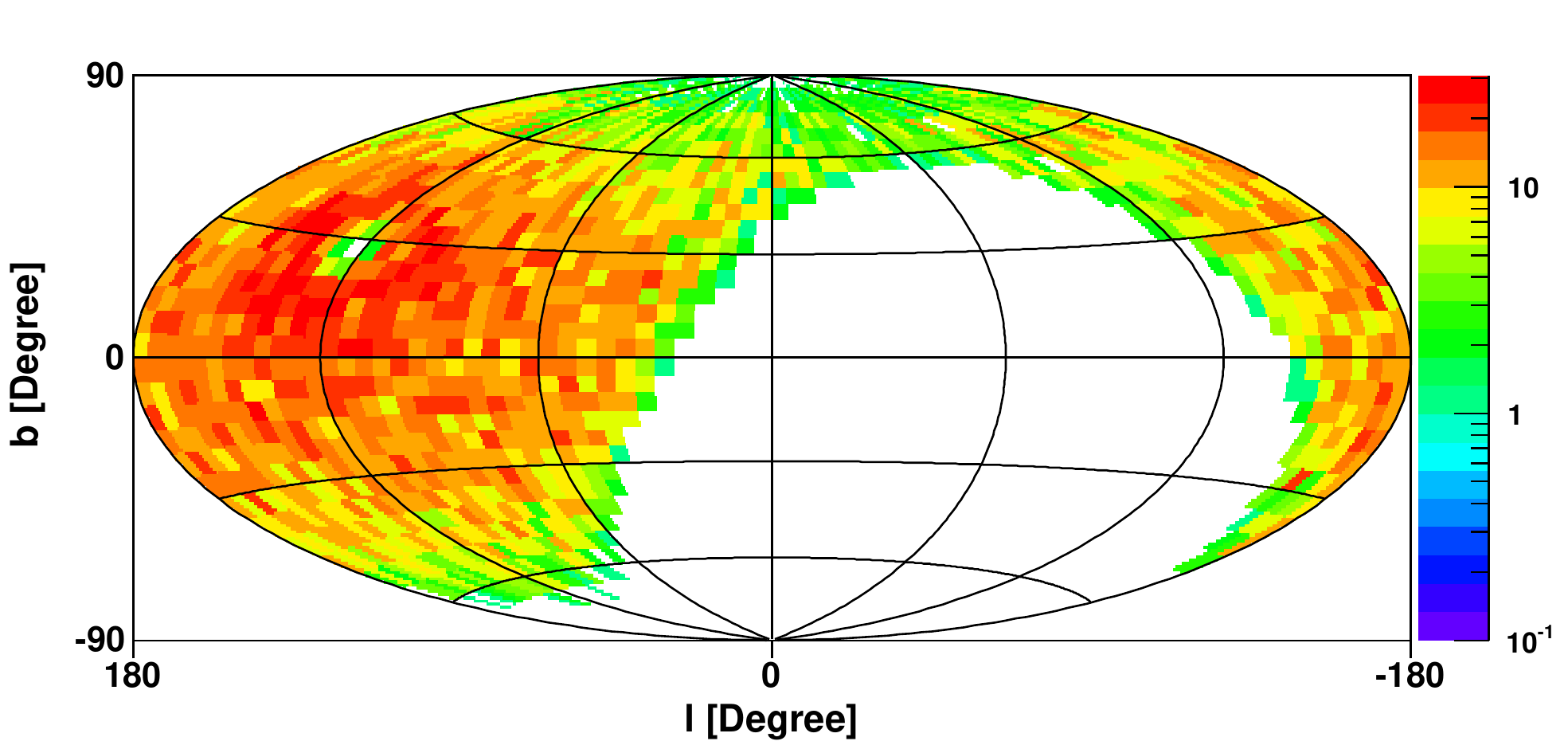}
\caption{Left: Distribution in Galactic coordinates of the expected
  number of events at the Telescope Array between 1 and 3~EeV in the
  case of a pure proton composition produced in stationary Galactic
  sources. Right: Distribution of the observed number of events at the
  Telescope Array between 1 and 3~EeV in Galactic latitude and
  longitude using $5^\circ\times5^\circ$ bins on the
  sky~\cite{TAAPP2016}.}
\label{fig:gal_protons_ta}
\end{figure}

Hence, the percent limits on the amplitude of the anisotropy exclude
the presence of a large fraction of Galactic protons at EeV
energies. Similar conclusions were recently obtained with data
recorded at the Telescope Array by fitting the angular distribution
observed at EeV energies to benchmark sky maps obtained by adding the
flux expected from stationary Galactic sources of protons on top of a
fraction of isotropic flux~\cite{TAAPP2016}. The exact modelling for
the expected flux of Galactic protons, taken from~\cite{Tinyakov2016},
is similar to the one described above. An example of benchmark sky map
for energies between 1 and 3~EeV is shown in the left panel of
Fig.~\ref{fig:gal_protons_ta}, while the sky map observed in this
energy range is shown in the right panel. At most, not more than
$\simeq 1\%$ of the observed flux can be described by the modelled sky
maps~\cite{TAAPP2016}.

Accounting for the inference from the depth of air shower maximum data 
from both the Pierre
Auger Observatory and the Telescope Array that protons are in fact abundant at
those energies~\cite{AugerPRD2014,Abbasi:2014sfa}, the lack of strong
anisotropies provides some indication that this component of protons is
extragalactic, gradually taking over a Galactic one. The low level of
anisotropy could then be the result of the addition of two vectors with opposite
directions, naturally reducing the amplitudes and producing the change of
phase observed at EeV energies. This scenario is to be explored with
additional data. Increased statistics is thus necessary to probe the
anisotropy contrast levels that may exist in this energy range and contain
valuable information about the old-standing question on the way the transition
between Galactic and extragalactic CRs occurs. Also, a current limitation of
the measurements is that neither spectra nor anisotropies can yet be studied
as a function of the mass of the particles with adequate statistical
precision, measurements that would allow a distinction between Galactic and
extragalactic angular distributions.

\subsection{Multipolar analysis above $\simeq 10~$EeV with full-sky coverage}

Above $\simeq 10~$EeV, the whole flux of UHECRs is expected to be of
extragalactic origin. Although the actual sources of UHECRs are still
to be identified, their distribution in the sky is expected to follow,
to some extent and as already stressed in previous sections, the
large-scale structure of the matter in the Universe. It is thus
interesting to highlight again that above 8~EeV, the amplitude shown
in Fig.~\ref{fig:1stharm} with a $p$-value as low as
$6.4\times10^{-5}$. Assuming that the only significant contribution to
the anisotropy is from a dipolar pattern, the amplitude of this signal
converts into a $(7.3\pm1.5)\%$ dipole
amplitude~\cite{AugerApJ2015b}. This hint may constitute in the near
future the first detectable signature of extragalactic CRs observed on
Earth.

To characterize further the angular distribution above 10~EeV, the
dipole moment on the sphere is of special interest. An unambiguous
measurement of this moment as well as of the full set of spherical
harmonic coefficients requires full-sky coverage. Currently, this can
be achieved piecemeal by combining data from observatories located in
both the northern and southern hemispheres. To this end, a joint
analysis using data recorded at the Pierre Auger Observatory and the
Telescope Array above $10$~EeV has been performed
in~\cite{AugerTA2014,Deligny2015}. Thanks to the full-sky coverage,
the measurement of the dipole moment reported in these studies does
not rely on any assumption on the underlying flux of CRs.

The main challenge in combining the data sets is to account adequately
for the relative exposures of both experiments. A band of declinations
around the equatorial plane is exposed to the fields of view of both
experiments, namely for declinations between $-15^\circ$ and
$25^\circ$. This overlapping region has been used for designing an
empirical procedure to get a relevant estimate of the relative
exposures: for an isotropic flux, the integrated energy spectra
measured independently by both experiments in the common band would
have to be identical. The commonly covered declination band could thus
be used for cross-calibrating empirically the energy spectra of the
experiments and for delivering an overall estimate of the relative
exposures. Since the shapes of the exposure functions are not
identical in the overlapping region, the observed energy spectra are
not expected to be identical in case of anisotropies. For small
anisotropies however, this guiding idea can nevertheless be
implemented in an iterative algorithm delivering finally estimates of
the relative exposures and of the multipole coefficients at the same
time. The uncertainties on the recovered coefficients, however, are
larger than expected from equation~\ref{eqn:rms_alm_bis} due to the
effect of the uncertainty in the relative exposures of the two
experiments. This propagation of uncertainty mainly impacts the
resolution in the dipole coefficient related to variations of the flux
in declination.

\begin{figure}[!ht]
  \centering
  \includegraphics[width=4.0in]{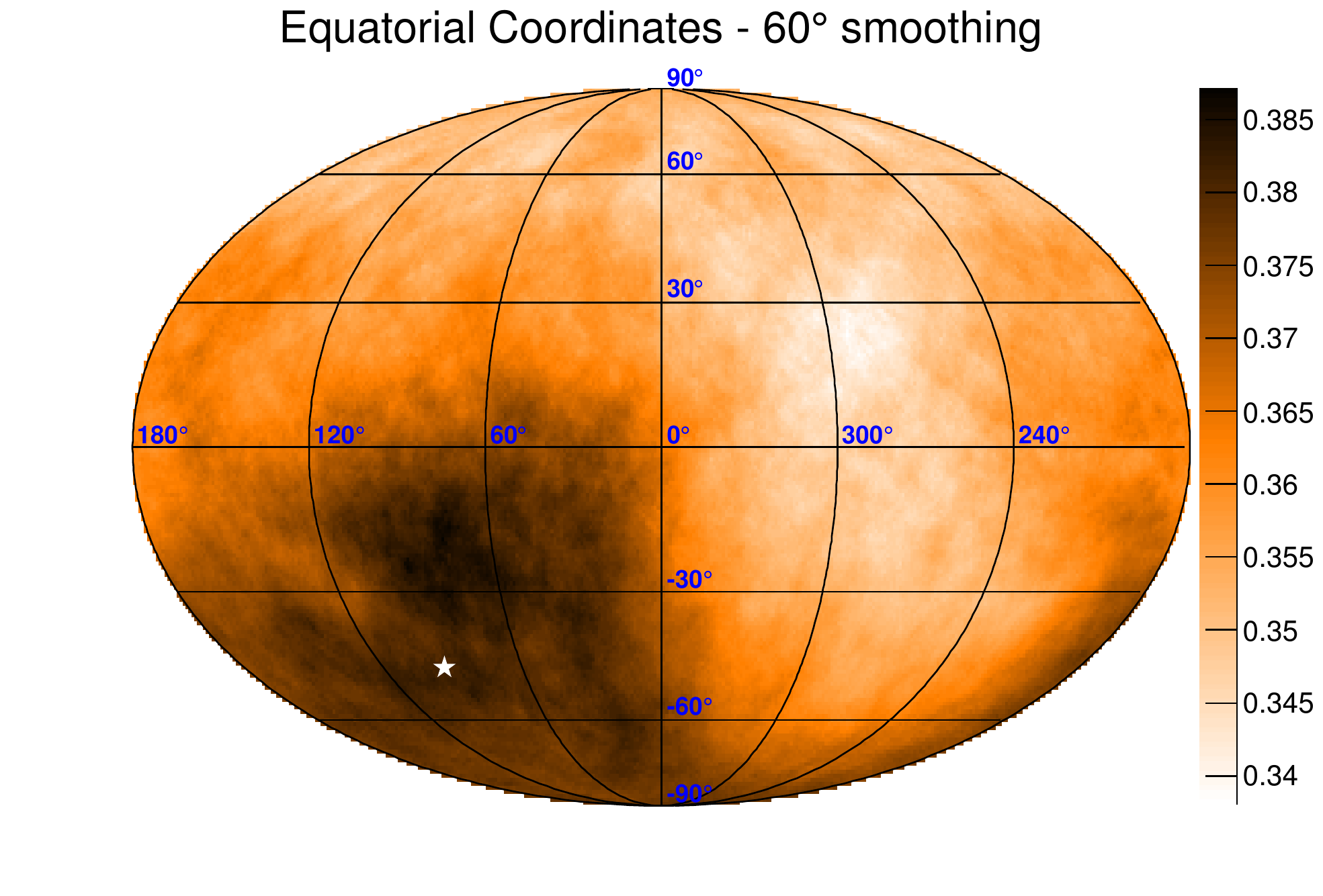}
  \caption{Sky map in equatorial coordinates of the average flux reconstructed
    from data recorded at the Pierre Auger Observatory and the Telescope Array
    above $10~$EeV smoothed out at a 60$^\circ$ angular scale, in
    km$^{-2}$~yr$^{-1}$~sr$^{-1}$ units~\cite{Deligny2015}.}
  \label{fig:skymap_augerta}
\end{figure}

The resulting entire mapping of the celestial sphere has revealed a
dipole moment with an amplitude $r=(6.5\pm 1.9)\%$, captured with a
chance probability of $5\times10^{-3}$~\cite{Deligny2015}. No other
deviation from isotropy has been observed at smaller angular scales. The
recovered moment can be visualized in Fig.~\ref{fig:skymap_augerta},
where the average flux smoothed out at an angular scale of $60^\circ$ per
solid angle unit is displayed using the Mollweide projection, in
km$^{-2}$~yr$^{-1}$~sr$^{-1}$ units. This map is drawn in equatorial
coordinates. The direction of the reconstructed dipole is shown as the
white star.

Large-scale anisotropies of CRs with energies in excess of $10~$EeV
are closely connected to the sources and the propagation mode of
extragalactic UHECRs, see \textit{e.g.}~\cite{Harari2014,Tinyakov2015}. Due to
scattering in the extragalactic magnetic fields, large deflections are
expected even at such high energies for field amplitudes in the nanogauss 
range and extended over coherence lengths of the order of one
megaparsec, or even for lower amplitudes if the electric charge of
UHECRs is large. For sources distributed in a way similar to the
matter in the Universe, the angular distribution of UHECRs is then
expected to be influenced by the contribution of nearby sources, so
that the Milky Way should be embedded into a density gradient of CRs
that should lead to at least a dipole moment. The contribution of
nearby sources is even expected to become dominant as the energy of
CRs increases due to the reduction of the horizon of UHECRs induced by
energy losses that are more important at higher energies. Once folded through
the Galactic magnetic field, the dipole pattern expected from this
mechanism is transformed into a more complex structure presumably
described by a lower dipole amplitude and higher-order
multipoles. However, in these scenarios, the dipole moment could
remain the only one within reach given the sensitivity of the current
generation of experiments. On the other hand, the detection of
significant multipole moments beyond the dipole could be
suggestive of non-diffusive propagation of UHECRs from sources
distributed in a non-isotropic way.

\section{Conclusions and outlook}
\label{sec:conclusions-outlook}

At ultra-high energies, the popular tropism that cosmic-ray astronomy may be
feasible relies on two assumptions: on the possibility that the messengers
have high enough rigidities to beat the magnetic blurring, and that the
reduced particle horizons caused by energy losses eliminate or suppress the
isotropic background from distant sources. However, for the same reason, the
pattern recognition of the astrophysical sites harboring the accelerators
which could be performed from UHECR arrival directions is made difficult by
the very small intensity of these particles. It is also possible that the
rigidity of the particles is not sufficient: properties of intervening
magnetic fields are uncertain, and the mass composition of UHECRs in the
highest-energy region remains largely unknown.

Thorough searches for various possible anisotropies of UHECRs conducted by the
Pierre Auger and Telescope Array collaborations during the past decade have
been summarized in this review. The data show a remarkable degree of isotropy
at all energies, with only a few hints on a possible anisotropic distribution,
like large-scale patterns in the EeV energy range, and the concentration of events 
(the hot spot) in the Northern hemisphere at the highest energies $E>57$~EeV 
(TA energy scale). The large-scale pattern above $\simeq 8~$EeV (Auger energy
scale) is best revealed through the first harmonic analysis in right ascension of
the Auger data. These indications still require confirmation with larger statistics.

With the data above $\simeq 55$~EeV released by both collaborations, it is worth
mentioning that other explorations of possible indications of anisotropies
were performed by numerous authors outside of these collaborations which were
not covered in this review, \textit{e.g.} searches for clustering of
events~\cite{CuocoXXX}; searches for correlations with extragalactic
matter~\cite{WaxmanXXX}; studies of the lensing effect of the Galactic
magnetic field to probe whether the observed arrival directions could
correlate with extragalactic objects once unfolded~\cite{FarrarXXX}; and
others.  Overall, although a few intriguing correlations with positions in the
sky tracing special astrophysical environments have been uncovered, none of
the corresponding significances is large enough to provide a compelling signal
of anisotropy at present, especially in view of the multiple searches
conducted.

The apparent isotropy can be explained by several factors. One of them is the
possibility that UHECRs are composed of a mix of light and intermediate/heavy
nuclei. In addition, the magnetic field in the halo of the Galaxy may be
underestimated due to the low density of electrons in these regions, so that
even for protons, the magnetic deflections could still be large at the highest
energies. The extragalactic fields, though bound by $\sim$~nG at large scales,
may be larger in the $\sim$~Mpc vicinity of our Galaxy in case it is embedded in a
filament. 

The absence of obvious bright sources on the UHECR sky, if not due to large
magnetic deflections, may indicate that the sources are too numerous and
individually weak.  This can be quantified by setting lower bounds on the
density of sources. Roughly, the absence of repeaters implies that the number
of contributing sources has to be larger than the square of the number of
events \cite{Dubovsky:2000gv}. Using the events above 70~EeV detected at the
Auger Observatory, the best up-to-date derived bounds, whose validity domains
are limited for magnetic deflections smaller than the angular scale $\psi$
used to search for the repeaters, range from $7\times10^{-4}~$Mpc$^{-3}$ at
$\psi=3^\circ$ up to $2\times10^{-5}~$Mpc$^{-3}$ at
$\psi=30^\circ$~\cite{AugerJCAP2013}.  For comparison, the density of
galaxies is roughly $10^{-2}~$Mpc$^{-3}$.

First attempts to get at full-sky surveys at ultra-high energies are being
developed between the two collaborations. The full-sky coverage is obviously
advantageous to probe all possible sources of UHECRs, and is indispensable for
the harmonic analysis aimed at revealing possible anisotropies at large
angular scales. Also, a multi-messenger approach could give clues for
deciphering the origin of the cosmic ray particles. In this context, a
full-sky study conducted in a collaboration among Auger, Telescope Array and
IceCube, has recently reported on the search for correlations between UHECRs
and very high-energy neutrino candidates detected by IceCube~\cite{AugerTAIC}.
It is interesting that the smallest post-trial $p$-values (corresponding to
significances slightly greater than $\simeq 3\sigma$) are obtained when
considering the correlations between the directions of cascade events observed
by IceCube and those of the UHECRs on an angular scale of $20^\circ$.  With
increased statistics, this kind of meta-analysis will help to understand
whether or not a contribution to the neutrino signal observed by IceCube
arises from the sources of the observed UHECRs.

Present detector exposures are already providing us with important constraints
on the origin of UHECRs. Even larger exposures will lead to much better
constraints and hopefully will eventually make possible the detection of the
brightest sources. The Telescope Array collaboration is building an extension
of the surface detector array, reaching a detection surface of about
$3000~$km$^2$. This will boost the UHECR statistics in the Northern hemisphere
which, apart from the sensitivity to sources in the Northern sky, is crucial
for the all-sky surveys. At the Auger Observatory, an upgraded instrumentation
is being deployed to equip the detectors with an additional plane of 4~m$^2$
of plastic scintillators above each station. This will provide us with
additional high-statistics measurements of the showers, helping to produce
composition-sensitive observables in an energy range including the highest
energies, and will open a possibility to perform mass-discriminated
anisotropy searches if the composition is mixed with light and heavy
elements~\cite{AugerPrime2015}. The operation of such an upgraded array is
anticipated between 2018 and 2024.

\section*{Acknowledgment}

We warmly thank Armando Di Matteo for providing us
Fig.~\ref{fig:energy_loss_length}.  The work of PT is supported in part by the
IISN project No. 4.4502.16 and Belgian Science Policy under IUAP VII/37 (ULB).


\begin{thebibliography}{99}
% intro
\bibitem{LinsleyPRL1963} J. Linsley, Phys. Rev. Lett. {\bf 10}, 146 (1963)
\bibitem{z-burst} D. Fargion, B. Mele \& A. Salis, Astrophys. J. {\bf 517}, 725 (1999)
\bibitem{AugerPRD2015_AugerJCAP2016_TAPRD2013} The Pierre Auger Collaboration, Phys. Rev. {\bf D91}, 092008 (2015); The Pierre Auger Collaboration, submitted to JCAP [arXiv:1612.01517]; T. Abu-Zayyad {\it et al.} [Telescope Array Collaboration], Phys. Rev. {\bf D88}, 112005 (2013)
\bibitem{AugerNIM2015} The Pierre Auger Collaboration, Nucl. Instrum. Meth. A {\bf 798}, 172 (2015)
\bibitem{TANIM2012} T. Abu-Zayyad \textit{et al.} [Telescope Array Collaboration] Nucl. Instrum. Meth. A {\bf 689}, 87 (2012)

% section 2.1
\bibitem{Aloisio:2006wv} R.~Aloisio {\it et al.}, Astropart. Phys. {\bf 27}, 76 (2007)   % [astro-ph/0608219].
\bibitem{GZK} K. Greisen, Phys. Rev. Lett. {\bf 16}, 748 (1966); G. Zatsepin \& V. Kuzmin, J. Exp. Theor. Phys. Lett. {\bf 4}, 78 (1966)
\bibitem{GZKprotons} C.~Hill \& D.~Schramm, Phys. Rev. {\bf D31}, 564 (1985); V.~Berezinsky \& S.~Grigorieva, Astron. Astrophys. {\bf 199}, 1 (1988); V.~Berezinsky, S.~Grigorieva \& V. A.~Dogiel, Astron. Astrophys. {\bf 232}, 583 (1990) 
\bibitem{GZKnuclei} J. L. Puget, F. Stecker \& J. H. Bredekamp,  Astrophys. J. {\bf 295}, 638 (1976); 
\bibitem{codes} E. Armengaud {\it et al.}, Astropart. Phys. {\bf 28}, 463 (2007); K.-H.~Kampert {\it et al.}, Astropart. Phys. {\bf 42}, 41 (2013); R.~Aloisio {\it et al.}, JCAP {\bf 1210}, 007 (2012); M.~De Domenico EPJP {\bf 128}, 99 (2013); O.~Kalashev \& E.~Kido, J. Exp. Theor. Phys. {\bf 120 5}, 790 (2015)
\bibitem{CRPropa3} R. Alves Batista \textit{et al.}, in prepatation
\bibitem{Gilmore2012} R. Gilmore \textit{et al.}, Mon.\ Not.\ Roy.\ Astron.\ Soc.\  {\bf 422}, 3189 (2012)

% section 2.2
\bibitem{SOPHIA} A. M\"ucke {\it et al.}, Computer Physics Communications, {\bf 124 2}, 290 (2000)
\bibitem{AlvesBatista2016} R. Alves Batista \textit{et al.}, JCAP {\bf 10}, 1510 (2015)
\bibitem{Kronberg1994_Pshirkov2016} P. P. Kronberg  Rep. Prog. Phys. {\bf 57}, 3
25 (1994); M.~S.~Pshirkov \textit{et al.}, Phys. Rev. Lett. {\bf 116}, 032007 (2016)
\bibitem{Dolag2005} K. Dolag \textit{et al.}, JCAP {\bf 0501}, 009 (2005)
\bibitem{Pshirkov2011} M. S. Pshirkov \textit{et al.}, Astrophys. J {\bf 738}, 192 (2011)
\bibitem{Farrar2012} G. Farrar and R. Jansson, Astrophys. J {\bf 757}, 14 (2012)
\bibitem{Tinyakov:2004pw} P.~G.~Tinyakov \& I.~I.~Tkachev, Astropart. Phys. {\bf 24}, 32 (2005) % [astro-ph/0411669].
\bibitem{Pshirkov:2013wka} M.~S.~Pshirkov, P.~G.~Tinyakov \& F.~R.~Urban, Mon.\ Not.\ Roy.\ Astron.\ Soc.\  {\bf 436}, 2326 (2013) % [arXiv:1304.3217 [astro-ph.CO]].
 
% section 3.1
\bibitem{LiMa1983} T.-P. Li \& Y.-Q. Ma, Astrophys. J. {\bf 272}, 317 (1983)
\bibitem{AugerApJ2015a} The Pierre Auger Collaboration, Astrophys. J {\bf 804}, 15 (2015)
\bibitem{TAApJ2014} R. U. Abbasi \textit{et al.} [Telescope Array Collaboration], Astrophys. J. {\bf 790}, L21 (2014)
\bibitem{AugerScience2007} The Pierre Auger Collaboration, Science {\bf 318}, 938 (2007)
\bibitem{Kawata:2015whq} K.~Kawata {\it et al.} [Telescope Array Collaboration], PoS ICRC {\bf 2015}, 276 (2016)  %%CITATION = POSCI,ICRC2015,276;%%
\bibitem{Sagawa:2015sgk} P.~Tinyakov {\it et al.} [Telescope Array Collaboration], PoS ICRC {\bf 2015}, 326 (2016)   %%CITATION = POSCI,ICRC2015,326;%%

% section 4.1
\bibitem{Huchra:2011ii} J.~P.~Huchra {\it et al.},  Astrophys.\ J.\ Suppl.\  {\bf 199}, 26 (2012) %  doi:10.1088/0067-0049/199/2/26  [arXiv:1108.0669 [astro-ph.CO]].  %%CITATION = doi:10.1088/0067-0049/199/2/26;%%
\bibitem{Baumgartner:2012qx} W.~H.~Baumgartner {\it et al.}, Astrophys.\ J.\ Suppl.\  {\bf 207} (2013) 19 %  doi:10.1088/0067-0049/207/2/19  [arXiv:1212.3336 [astro-ph.HE]].  %%CITATION = doi:10.1088/0067-0049/207/2/19;%%
\bibitem{vanVelzen:2012fn} S.~van Velzen {\it et al.}, Astron.\ Astrophys.\  {\bf 544}, A18 (2012) %  doi:10.1051/0004-6361/201219389  [arXiv:representatio1206.0031 [astro-ph.CO]].
  %%CITATION = doi:10.1051/0004-6361/201219389;%%
\bibitem{Abu-Zayyad:2013vza} T.~Abu-Zayyad {\it et al.} [Telescope Array Collaboration], Astrophys.\ J.\  {\bf 777}, 88 (2013) %  doi:10.1088/0004-637X/777/2/88   [arXiv:1306.5808 [astro-ph.HE]].  %%CITATION = doi:10.1088/0004-637X/777/2/88;%%
\bibitem{Laing:1983ke} R.~A.~Laing, J.~M.~Riley \& M.~S.~Longair, Mon.\ Not.\ Roy.\ Astron.\ Soc.\  {\bf 204}, 151 (1983)  %%CITATION = MNRAA,204,151;%%
\bibitem{Fermi-LAT:2011xmf} Fermi-LAT Collaboration, Astrophys.\ J.\  {\bf 743}, 171 (2011)%  doi:10.1088/0004-637X/743/2/171  [arXiv:1108.1420 [astro-ph.HE]].
  %%CITATION = doi:10.1088/0004-637X/743/2/171;%%
\bibitem{VeronCetty:2006zz} M.-P.~Veron-Cetty \& P.~Veron, Astron.\ Astrophys.\  {\bf 455}, 773 (2006)%  doi:10.1051/0004-6361:20065177  %%CITATION = doi:10.1051/0004-6361:20065177;%%

% section 4.2
\bibitem{AbuZayyad:2012hv} T.~Abu-Zayyad {\it et al.} [Telescope Array Collaboration], Astrophys.\ J.\  {\bf 757}, 26  (2012)%  doi:10.1088/0004-637X/757/1/26   [arXiv:1205.5984 [astro-ph.HE]].  %%CITATION = doi:10.1088/0004-637X/757/1/26;%%
\bibitem{Abbasi:2014sfa} R.~U.~Abbasi {\it et al.} [Telescope Array Collaboration], Astropart.\ Phys.\  {\bf 64}, 49 (2015) %  doi:10.1016/j.astropartphys.2014.11.004  [arXiv:1408.1726 [astro-ph.HE]].  %%CITATION = doi:10.1016/j.astropartphys.2014.11.004;%%
\bibitem{Aab:2014aea} The Pierre Auger Collaboration, Phys.\ Rev.\ D {\bf 90}, 122006 (2014) %  doi:10.1103/PhysRevD.90.122006  [arXiv:1409.5083 [astro-ph.HE]].
  %%CITATION = doi:10.1103/PhysRevD.90.122006;%%
\bibitem{Koers:2009pd} H.~B.~J.~Koers \& P.~Tinyakov, Mon.\ Not.\ Roy.\ Astron.\ Soc.\  {\bf 399}, 1005 (2009) %  doi:10.1111/j.1365-2966.2009.15344.x  [arXiv:0907.0121 [astro-ph.CO]].  %%CITATION = doi:10.1111/j.1365-2966.2009.15344.x;%%
\bibitem{Koers:2008ba}  H.~B.~J.~Koers and P.~Tinyakov,
  %``Testing large-scale (an)isotropy of ultra-high energy cosmic rays,''
  JCAP {\bf 0904} (2009) 003
    %%CITATION = doi:10.1088/1475-7516/2009/04/003;%%

% section 5
\bibitem{Linsley1975} J. Linsley, Phys. Rev. Lett. {\bf 34}, 1530 (1975)
\bibitem{Deligny2013} O. Deligny \& F. Salamida, Astropart. Phys. {\bf 46}, 40 (2013)
\bibitem{Sommers2001} P. Sommers, Astropart. Phys. {\bf 14}, 271 (2001)
\bibitem{Billoir2008} P. Billoir \& O. Deligny, JCAP {\bf 02}, 009 (2008)
\bibitem{AlSamarai2015} I. Al Samarai for the Pierre Auger Collaboration, PoS ICRC {\bf 2015}, 372 (2016) 
\bibitem{UHECR2012} P. Tinyakov {\it et al.}, EPJ Web of Conferences {\bf 53}, 01008 (2013)
\bibitem{Yakutsk2001} V. P. Egorova {\it et al.}, Journ. Phys. Soc. Japan, Suppl. B \textbf{70} (2001); M. I. Pravdin {\it et al.}, J. Exp. Theor. Phys. \textbf{92}, 766 (2001)
\bibitem{Linsley1963} J. Linsley, Proceedings of the 8th ICRC, Jaipur, {\bf 4}, 77 (1963)
\bibitem{DeAlmeida2013} R. De Almeida for the Pierre Auger Collaboration, Proc. 33rd ICRC, Rio de Janeiro, Brazil, 2013
\bibitem{AugerApJ2015b} The Pierre Auger Collaboration, Astrophys. J {\bf 802}, 111 (2015)
\bibitem{Edge1978} D. Edge {\it et al.}, J. Phys. G {\bf 4}, 133 (1978)
\bibitem{AugerApJS2012} The Pierre Auger Collaboration, Astrophys. J. Suppl. {\bf 203}, 34 (2012)
\bibitem{TAAPP2016} The Telescope Array Collaboration, Astropart. Phys. {\bf 86}, 21 (2017)
\bibitem{Tinyakov2016} P. Tinyakov {\it et al.}, Mon. Not. Roy. Astron. Soc. {\bf 460}, 3479 (2013)
\bibitem{AugerPRD2014} The Pierre Auger Collaboration, Phys. Rev. {\bf D90}, 122005 (2014)
\bibitem{AugerTA2014} The Pierre Auger and Telescope Array Collaborations, Astrophys. J {\bf 794}, 172 (2014) 
\bibitem{Deligny2015} O. Deligny for the Pierre Auger and Telescope Array Collaborations, PoS ICRC {\bf 2015}, 395 (2016)
\bibitem{Harari2014} D. Harari, S. Mollerach \& E. Roulet, Phys. Rev. {\bf D89}, 123001 (2014)
\bibitem{Tinyakov2015} P. Tinyakov \& F. Urban, J. Exp. Theor. Phys. {\bf 120}, 3, 533,  (2015)

% conclusion
\bibitem{CuocoXXX} A. Cuoco {\it et al.}, Astrophys. J. {\bf 702}, 825 (2009); H.~Takami \& K.~Sato, Astropart. Phys. {\bf30}, 306 (2009); H.~Takami, K.~Murase \& C.~D.~Dermer, Astrophys. J {\bf 817}, 59 (2016)
\bibitem{WaxmanXXX} T. Kashti \& E. Waxman, JCAP {\bf 05}, 006 (2008); D. Gorbunov {\it et al.}, J. Exp. Theor. Phys. Lett. {\bf 87}, 461 (2008); D.~Fargion, Physica Scripta {\bf 78}, 045901 (2008); M.~R.~George {\it et al.}, Mon.\ Not.\ Roy.\ Astron.\ Soc.\  {\bf 388}, L59 (2008); T.~Wibig \& A.~Wolfendale, The Open Astronomy Journal {\bf 2}, 95 (2009);  I.~V.~Moskalenko {\it et al.}, Astrophys. J {\bf 693}, 1261 (2009); A.~Berlind, G.~Farrar \& I.~Zaw, Astrophys. J {\bf 716}, 914 (2010); L.~J.~Watson, D.~J.~Mortlock \& A.~H.~Jaffe, Mon.\ Not.\ Roy.\ Astron.\ Soc.\  {\bf 418}, 206 (2011); F.~Oikonomou {\it et al.}, JCAP {\bf 1305}, 015 (2013)
\bibitem{FarrarXXX} A.~Keivani, G.~R.~Farrar \& M.~Sutherland, Astropart. Phys. {\bf 61}, 47 (2015); N.~M.~Nagar \& J.~Matulich, Astron. Astrophys. {\bf 523}, A49 (2010)
\bibitem{Dubovsky:2000gv}  S.~L.~Dubovsky, P.~G.~Tinyakov \& I.~I.~Tkachev,
  %``Statistics of clustering of ultrahigh energy cosmic rays and the number of their sources,''
  Phys.\ Rev.\ Lett.\  {\bf 85}, 1154 (2000)
  %%CITATION = doi:10.1103/PhysRevLett.85.1154;%%
\bibitem{AugerJCAP2013} The Pierre Auger Collaboration, JCAP {\bf 05}, 009 (2013)
\bibitem{AugerTAIC} The Pierre Auger, Telescope Array and Ice Cube Collaborations, JCAP {\bf 01}, 037 (2016)
\bibitem{AugerPrime2015} The Pierre Auger Collaboration, arXiv:1604.03637




%\bibitem{Dol04} Dol04
%\bibitem{Fan14} Fan14
%\bibitem{He14} He14
%\bibitem{Abr10} Abr10
% section 3.2
\end{thebibliography}
\end{document}